\documentclass{article}

\usepackage{PRIMEarxiv}

\usepackage[utf8]{inputenc} % allow utf-8 input
\usepackage[T1]{fontenc}    % use 8-bit T1 fonts
\usepackage{hyperref}       % hyperlinks
\usepackage{url}            % simple URL typesetting
\usepackage{booktabs}       % professional-quality tables
\usepackage{amsfonts}       % blackboard math symbols
\usepackage{nicefrac}       % compact symbols for 1/2, etc.
\usepackage{lipsum}
\usepackage{fancyhdr}       % header
\usepackage{graphicx}       % graphics
\graphicspath{{media/}}     % organize your images and other figures under media/ folder

\usepackage[table]{xcolor}

\usepackage[numbers]{natbib}
\usepackage{algorithm}
\usepackage[noEnd=false]{algpseudocodex}

\usepackage{setspace}

\usepackage{paralist}

\usepackage{color}

\usepackage[all]{xy}
\usepackage{amsmath}

\usepackage[title,titletoc]{appendix}

\usepackage{tabularx}
\usepackage{longtable}
\usepackage{threeparttable}
\usepackage{multirow}

\usepackage{numprint}
\npstyleenglish

\usepackage{caption}
\usepackage{subcaption}

\usepackage{framed}
\usepackage{multicol}
\usepackage{nomencl}
\makenomenclature

\DeclareTextFontCommand{\emph}{\bfseries}

\usepackage{etoolbox}
\renewcommand\nomgroup[1]{%
  \item[\bfseries%\itshape
  \ifstrequal{#1}{A}{Acronyms}{%
  \ifstrequal{#1}{M}{Patient pathway model components}{%
  \ifstrequal{#1}{D}{Dissimilarity algorithm}{%
  \ifstrequal{#1}{S}{Mining method}{%
  \ifstrequal{#1}{O}{Other variables}{%
  }}}}}%
]}
\title{Framework based on complex networks to model and mine patient pathways
}

\author{
  Caroline de Oliveira Costa Souza Rosa\\
  Ph.D. candidate in Computational Modeling \\
  Laborat\'orio Nacional de Computa\c c\~ao Cient\'ifica (LNCC) \\
  Petr\'opolis\\
   \And
  M\'arcia Ito \\
  M.Sc. Program in Productive Systems \\
  Centro Paula Souza (PPG-GTPS) \\
  S\~ao Paulo\\
  \And
  Alex Borges Vieira \\
  Computer Science Department \\
  Universidade Federal de Juiz de Fora (UFJF) \\
  Juiz de Fora \\
  \And
  Klaus Wehmuth \\
  Laborat\'orio Nacional de Computa\c c\~ao Cient\'ifica (LNCC) \\
  Petr\'opolis \\
  \And
  Ant\^onio Tadeu Azevedo Gomes \\
  Laborat\'orio Nacional de Computa\c c\~ao Cient\'ifica (LNCC) \\
  Petr\'opolis \\
}

\begin{document}
\maketitle

\begin{abstract}
The automatic discovery of a model to represent the history of encounters of a group of patients with the healthcare system---the so-called ``pathway of patients''---is a new field of research that supports clinical and organisational decisions to improve the quality and efficiency of the treatment provided.
The pathways of patients with chronic conditions tend to vary significantly from one person to another, have repetitive tasks and demand the analysis of multiple perspectives (interventions, diagnoses, medical specialities, among others) influencing the results.
Therefore, modelling and mining those pathways is still a challenging task. 
In this work, we propose a framework comprising:
\begin{inparaenum}[(i)]
    \item a pathway model based on a multi-aspect graph, 
    %\item a novel dissimilarity measurement to compare pathways taking the elapsed time into account, and 
    \item a personalised dissimilarity measurement to compare pathways taking the elapsed time into account, and 
    \item a mining method based on traditional centrality measures to discover the most relevant steps of the  pathways.
\end{inparaenum} 
We evaluated the framework using the study cases of pregnancy and diabetes, which revealed the framework's usefulness in finding clusters of similar pathways, representing them in an easy-to-interpret way and highlighting the most significant patterns according to multiple perspectives.
\end{abstract}

% keywords can be removed

\section{Introduction}
\label{sec:intro}

The development of clinical guidelines and the health system planning are measures that optimise the treatment outcomes of patients while securing a sustainable care provision. Both of them rely on the notion that a patient navigates through the system, and the succession of their encounters must follow a timely and proper sequence of steps to maximise the probability of having good prognoses. 
From the clinical point of view, the treatment process of a patient should follow the state-of-art recommendations for a given condition. From the health service point of view, the expected treatment process of the patients determine the health units distribution and resources required from them to meet the patients' demands.
A task as important as designing and implementing such measures is evaluating whether patients are actually following them and if there are improvement opportunities. In other words, it is important to obtain a model that describes not the desired but the true patients' pathways. 

The assessment of patient pathways is not a new topic of interest~\cite{Rotondi1997}, but the growing electronic record of healthcare data and the development of computational techniques to analyse them have hastened the progress and expanded the possibilities of studying them. 
Besides retelling the sequence of interventions a patient underwent~\cite{Andrews2020,Gonzalez-Garcia2020,Kempa-Liehr2020}, the patient pathway may also reveal the disease evolution~\cite{DeOliveira2020b,Khan2018,Prodel2018}, the healthcare units/departments visited~\cite{Durojaiye2018,Arnolds2018,Rismanchian2017} or the specialities advising the patient's treatment~\cite{Conca2018,Senderovich2016}, depending on the perspective from which the patient pathway is defined~\cite{Manktelow2022,Dhaenens2018}. 
These various perspectives may interfere with each other, such as when studying the treatment of patients spread over a large geographic area with multiple healthcare units~\cite{Egho2014} or when the intervention, the purpose of the consultation, the medication and the evolution of the disease are all relevant to following patients with chronic diseases~\cite{Zhang2015}, demanding the construction of patient pathways with multiple perspectives.

A patient pathway model aims to represent what happens (the steps of the pathway according to the chosen perspective) and how it happens, e.g.\ when, in which order, or with which causal dependencies. The steps are usually represented as an item of a sequence~\cite{Antonelli2012,Aspland2021}, a node of a graph~\cite{Najjar2018,Basole2015} or an activity in a process model~\cite{Andrews2020,Kempa-Liehr2020}.
The task of mining patient pathways consists of automatically obtaining the pathway model from healthcare data. This involves identifying the most relevant encounters, direct successions, mutually exclusive patterns and concurrent tasks, among others. 
Although patients with the same health condition are likely to share treatment patterns, case studies frequently report a large number of different pathways when compared to the number of patients. Thus, one of the main challenges for doing so is dealing with the high \emph{variability} of patient pathways, that is, the number of distinct pathways.
Many factors contribute to such variety: 

\begin{itemize}
    \item Patient singularities. Different from product manufacturing, where process standardisation leads to uniform results, repeating the same treatment approach does not guarantee the same outcome for every patient. The reasons are diverse, including genetics, allergies or comorbidities that restrain first-line therapy and demand personalised treatment plans; 
    \item Influence of medical decisions. 
    Except for contexts where clear guidelines are available and rigidly implemented, the process of treating a patient has an \textit{ad hoc} nature~\cite{Rebuge2012}, and is influenced by factors such as the clinician experience and the healthcare unit culture;
    \item Influence of the patient's decisions. Especially in primary care, patients have a leading role in their pathways. They can decide whether or not they will follow the prescribed treatment or return for a follow-up when recommended. Moreover, for many illnesses, lifestyle choices have a massive impact on health outcomes~\cite{PatientSelfManagement}.
    %\item Area dynamism - new diseases, diagnostic methods, treatment options --> há um tópico exatamente igual no artigo do Rebuge; fiquei sem saber como trazer algo novo
\end{itemize}

The healthcare condition involved in the case study might exacerbate or mitigate those characteristics. For acute conditions, it is common to have an established clinical pathway and an expected outcome within a short/moderate time interval, which might contribute to lessening the number of pathway variants. On the other hand, the complexity of patient pathways tends to be larger for chronic conditions. Chronic health condition refers to a long-lasting state of one's health that demands continuous care and, possibly, episodes of acute care as well~\cite{Mendes2018}. They comprise not only chronic diseases, but also persistent infectious diseases, mental illnesses, and even health maintenance (e.g.\ follow-up of pregnant, child, elderly, or disabled patients), among others~\cite{Mendes2018}.
They are commonly associated with long pathways that span over months or even years. The increased pathway length and a likely larger number of different interventions, diagnoses, and specialities cooperate to an intensified pathway variability.

The treatment of chronic conditions usually demands some interventions to occur repetitively for the same patient, such as period screening consultations for post-cancer patients~\cite{Rinner2018}. Thus, the pathway model must provide the means to represent these repetitive tasks in an interpretable way. 
Another important feature for modelling and mining chronic condition pathways is preserving time intervals between the events. Although timely provision of care is vital in any scenario, the inclusion of time information in the pathway of patients with chronic conditions not only enriches the model but could also discriminate different behaviours~\cite{ZhangPadman2015}. 
Moreover, it is common for patients with chronic conditions to have other health issues, either due to multimorbidity or comorbidity. Hence, while designing the study the researchers might opt for a flexible mining method to discover pathway models comprising other health conditions~\cite{Najjar2018}---overlapping pathways---or focus on the main pathway while cleaning the data for the study~\cite{Zaballa2020}. Lastly, the mutual influence of different perspectives of a patient pathway becomes more apparent for chronic conditions, such as when the healthcare professionals coordinating the care could influence the outcomes of the treatment pathway~\cite{Conca2018}.

Despite the challenges for modelling and discovering patient pathways related to chronic conditions, their assessment is particularly useful because some of these conditions do not have clear clinical guidelines to direct the patients' treatment, so identifying pathways that led to positive or negative outcomes could support the development of the guidelines. Moreover, reducing deaths related to chronic diseases (non-communicable diseases) is one of the goals of the 2030 Agenda for Sustainable Development of the United Nations~\footnote{\url{https://sdgs.un.org/2030agenda}}, so understanding how engaged patients are with their treatment and which moments in their pathways could be improved are means of supporting public health policies to achieve this goal.

\subsection{Related Work}

% perspective ... usually represented as a point of a sequence, a node of a graph or an activity in a process model
% direct successions ....

%When the model is expressed as a sequence, the encounters are its items and the model depicts direct successions between them. However, a sequence can only display one patient pathway at a time, so, to study a group of individuals, a list of frequent patterns becomes necessary. Alternatively, a graph whose nodes are the encounters and the directed edges represent which followed one another in at least one pathway can summarise the behaviours of a group of patients in a single model, and even generalise it allowing some unobserved pathways. Similarly to sequence mining, many authors identified frequent nodes and edges to obtain a patient pathway graph.
%Although sequences and sequence mining methods were frequently used, process mining techniques have overtaken ,,, other methods are also used such as
% fuzzy miner -- repetição de rotulos

In the context of chronic conditions, there are several examples of models and mining methods to discover patient pathways in the literature.
\citeauthor{Egho2014} proposed a novel sequence mining method that fits sequences of multidimensional items or itemsets. For example, their case study comprised lung cancer patients whose pathway was defined as the sequence of hospitalisations. Each hospitalisation contained information on the healthcare unit, the diagnosis, and the medical procedures. The mining method considers different levels of granularity for each perspective and obtains the most specific subsequences whose support is greater than a given threshold. The authors later proposed a similarity measure for this model~\cite{Egho2015}.
A drawback of mining frequent subsequences is that, depending on the variability of pathways, the list of sequences for the specialists to evaluate may be too long, making it hard for them to draw inferences. 

Instead of listing sequences of frequent patterns, \citeauthor{Zhang2015} used a graph to represent a set of pathways. Their work involved chronic kidney disease patients who suffered at least one acute kidney injury. They modelled patient pathways considering four perspectives: type of visit, diagnoses, medical procedures, and medications. The authors concatenated these distinct perspectives to create a new label for each encounter (named ``super node''), as a means of absorbing multiple facets and dealing with the fact that some encounters involved multiple diagnoses and medications, due to comorbidities of the patients. Then, the authors created a Markov model whose states correspond to a pair of super nodes that happened immediately after each other in the pathways. The pathway mining method consisted of identifying the most frequent states and transitions in the Markov model and adding them to a directed graph.   

% stefanini 2020 novo -> uso de técnicas tradicionais de PM (foco em modelo de processo) para casos de alta variabilidade
% acho melhor deixar este para a discussão !!!!
%\citeauthor{Stefanini2020novo} tackles the problem of obtaining a model expressed as process algebra in a context of unstructured processes, such as their case study of lung cancer patient pathways in a hospital. Their method consists of  

% najjar
%the heart failure case study conducted by \citeauthor{Najjar2018} 

\citeauthor{Dagliati2017} also used a graph, but an acyclic-directed one---also called a directed tree graph---to represent patient pathways, so if an activity occurred more than once in a pathway, it would appear in different nodes to prevent the existence of cycles. This improves the legibility of the model, especially if some tasks occur repeatedly in the pathways.
The authors mined such a model adding only frequent behaviour to the graph, and enriched the edges with flow and time distribution information.
In a later work, the authors extended the method to identify concurrent activities~\cite{Dagliati2018}. 

A drawback of using acyclic graphs is that, unless the variability is small, the model might become too large and hard to interpret. Moreover, it is a model that overfits data.
As an alternative, from an initial acyclic graph representing all pathways, \citeauthor{Duma2020} developed a mining method that identifies subgraphs (``branches'' of the tree graph) that are not frequent enough according to a specified threshold and converts them into a cyclic graph. They named the resulting model as Hybrid Activity Tree.

Another alternative is the Time Grid Process Model proposed by \citeauthor{DeOliveira2020a}. It contains time layers that indicate in which position of the pathway an activity occurred. Therefore, an activity can appear in different layers of the model. This construction prevents the origination of cycles and improves the model readability. As an activity can appear only once per layer, the model does not grow as much as a tree graph and its generalisation is also improved.
The authors cluster the time intervals associated with each edge to identify how many time patterns exist---if all patients whose pathway contains the edge executed it in a similar interval, there will be a single edge, but if different behaviours are identified, multiple edges are added.  
Such a model is obtained from data using an optimisation algorithm with the goal of maximising the capability of the model to reproduce the pathways in the event log.  

Despite the advancements in the related area, there is still a need for mining methods and legible models that support the assessment of pathways with repetitive tasks from a multi-perspective approach, focusing on identifying not only frequent patterns but also relevant ones, such as treatment complications, which may not be common but that are worthy of attention.

% \subsection{Patient pathway modelling}
% \label{sec:relatedWork_model}
% \subsection{Patient pathway mining}
% \label{sec:relatedWork_mining}

\subsection{Research Goals} 
The objective of this work is multi-fold. First, we aim to contribute with a new patient pathway model that enables studying pathways from multiple perspectives while providing a good representation of pathways with repetitive activities. 
Second, we propose a personalised dissimilarity measurement which suits the comparison of patient pathways and enables clustering them.
Finally, we develop a mining method based on this model to automatically discover the pathway model from healthcare data. 
To attain the first objective, we present a model built as a MultiAspect Graph (MAG), which is a generalisation of the traditional graph that supports the representation of time-varying and/or multi-layer networks~\cite{Wehmuth2016}. 
As regards the second goal, we present as algorithm specially designed to compare sequences of items with time intervals between them.
We deal with the third objective by using traditional centrality measures to simplify the MAG and obtain the pathway model.

The remainder of this paper is organised as follows: Section~\ref{sec:methods} introduces the framework for modelling and mining patient pathways, which includes the pathway model, a novel dissimilarity measurement used for clustering pathways, and the mining method. It also describes the two real case studies we selected to test the framework. In Section~\ref{sec:results}, we present the results of our analyses and in Section~\ref{sec:conclusion}, we discuss the results and conclude the article. 

% ----------------------------------------------
\section{Methods}
\label{sec:methods}

This section will start with a brief presentation of some of the terminology this work adopts~(Section~\ref{subsec:terminology}). Next, in Section~\ref{subsec:mag_model}, we introduce our patient pathway model; in Section~\ref{subsec:data}, we present the dataset used in the study cases; in Section~\ref{subsec:similarity} we propose a similarity measure to compare and cluster the pathways; we describe the clustering step in Section~\ref{subsec:clustering}; finally, in Section~\ref{subsec:simplification} we introduce a pathway mining method compatible with the proposed model.

%The remainder of this section is organised as follows: in Section~\ref{subsec:mag_model}, we introduce our patient pathway model; in Section~\ref{subsec:data}, we present the dataset used in the study cases; in Section~\ref{subsec:similarity} we propose a similarity measure to compare and cluster the pathways; finally, in Section~\ref{subsec:simplification} we introduce a process mining method compatible with the proposed pathway model.

\subsection{Basic terminology}
\label{subsec:terminology}
%Our approach to modelling and mining patient pathways is analogous to the sub-field of Process Mining that deals with mining unstructured process models~\cite{Gunther2007,Stefanini2020novo}, but focusing on multi-perspective analysis. --> passei este trecho para a discussão
%An activity tuple is a tuple that contains information about the different perspectives from which a ``step'' in the patient's pathway could be defined. 
The history of encounters of a patient with the healthcare system is a patient pathway. 
An \emph{activity tuple} represents each step of the pathway and each element of the tuple expresses one \emph{perspective} from which we can scan the pathway.
For instance, in related work, authors have modelled patient pathways as sequences of medical interventions~\cite{Andrews2020,Sato2020},diagnoses~\cite{DeOliveira2020b,Khan2018}, hospital departments~\cite{Durojaiye2018,Arnolds2018}, or medical specialities~\cite{Conca2018}. 
Using an activity tuple instead of a single activity allows for two or more of these perspectives to be considered concomitantly.  
The record of an activity tuple for a specific patient is called an \emph{event}.
To apply our approach to a certain data set, it must contain timestamped events, each one associated with a patient and an activity tuple. 

%The remainder of this section is organised as follows: in Section~\ref{subsec:mag_model}, we introduce our patient pathway model; in Section~\ref{subsec:data}, we present the dataset used in the study cases; in Section~\ref{subsec:similarity} we propose a similarity measure to compare and cluster the pathways; finally, in Section~\ref{subsec:simplification} we introduce a process mining method compatible with the proposed pathway model.

% 'complete' mag
\subsection{Patient Pathways as a MultiAspect Graph}
\label{subsec:mag_model}

The \textit{MultiAspect} Graph~(MAG)~\cite{Wehmuth2016} is a generalisation of the graph concept, meant to model the relationship between objects in time-varying and/or multi-layer networks. 
A traditional graph is defined as a set of nodes and a set of edges, while a MAG has a set of aspects and a set of edges.
An \emph{aspect} represents a feature of the modelled network. 
Instead of nodes that typify a single entity that cannot be subdivided, the nodes of a MAG are $n$-tuples, whose length $n$ is the number of aspects of the MAG. 

In the patient pathway modelling domain---especially in the context of chronic illness, where the same medical procedure can appear repeatedly in the patient pathway---it is important to identify relevant nodes differentiating the context (layer) and the moment (time) it occurred. 
Thus, the MAG emerges as a feasible option for modelling such pathways. 

In this work, besides the sequence of medical interventions a patient underwent, we are interested in the professionals (occupations) who performed the interventions and the healthcare units where they occurred.  
Therefore, we propose the MAG $H = (I,O,U,S)$, where $I$ is the aspect Intervention, $O$ is the aspect Occupation, $U$ is the aspect Healthcare Unit, and $S$ is the aspect Sequence, which keeps the order of the events in the patient's pathway.

\begin{figure}[!ht]
    \centering
    \includegraphics[width=0.9\textwidth]{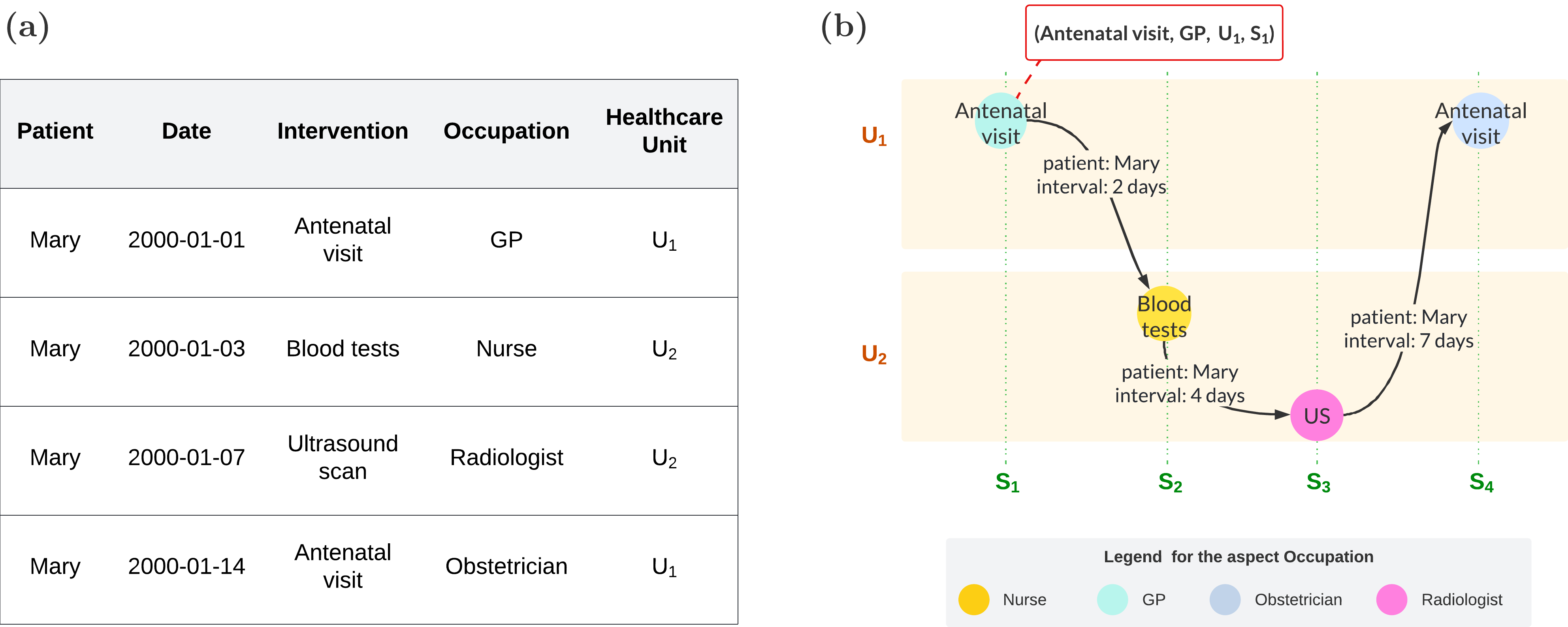}
    \caption{Example of converting an (a)~event log containing a patient pathway to (b)~a pictorial representation of the corresponding MAG, as proposed in this work. The aspect Intervention has $3$ elements (Antenatal visit, Blood tests and Ultrasound scans); the aspect Occupation has $4$ elements (GP, Nurse, Radiologist, and Obstetrician); the aspect Healthcare Unit has $2$ elements ($U_1$ and $U_2$); and the aspect Sequence has $4$ elements ($S_1$-$S_4$). Each circle represents a node (activity tuple); their colour indicate the Occupation; their labels indicate the Intervention; their horizontal position (following the dashed vertical lines) indicates the Sequence; and the stacked horizontal bar where each node is located represents the Healthcare Unit.}
    \label{fig:example-mag}
\end{figure}

Figure~\ref{fig:example-mag}a shows a fictitious patient pathway described as an event log.
The first event of the pathway is characterised by the   intervention ``Antenatal visit'', the occupation ``GP'' and the healthcare unit ``$U_1$''. 
Thus, its corresponding activity tuple is \textit{(Antenatal visit, GP, $U_1$, $S_1$)}, where $S_1$ is the element of the aspect Sequence that denotes that the activity tuple is linked to the first event of the pathway. 
Figure~\ref{fig:example-mag}b shows a pictorial representation of the MAG corresponding to this same patient pathway.%: the stacked horizontal bars represent the aspect Healthcare Unit ($U_1$ and $U_2$); the vertical lines represent the aspect Sequence ($S_1$ to $S_4$); the coloured circles represent the aspect Occupation; and the text within each coloured circle represents the aspect Intervention.
The first event of the pathway is represented by the upper leftmost circle in Figure~\ref{fig:example-mag}b.
%Following the table in the left, the first event of the patient pathway is represented as the activity tuple \textit{(Antenatal visit, GP, $U_1$, $S_1$)}, where $S_1$ is the element of the aspect Sequence which indicates that the remaining elements of the tuple (\textit{(Antenatal visit, GP, $U_1$)}) happened in the first event of Mary's pathway. 

The edges of a MAG are tuples of length $2n$, where $n$ is the number of aspects of the model. The first $n$ elements correspond to the origin node, while the last $n$ elements correspond to the destination node. 
For example, the edge that leaves the node corresponding to the upper leftmost circle depicted in Figure~\ref{fig:example-mag}b is \textit{(Antenatal visit, GP, $U_1$, $S_1$, Blood tests, Nurse, $U_2$, $S_2$)}.
%The exemplified pathway has four events and, as there is only one patient, there are four nodes in the MAG visualisation.  
Moreover, the edges of the proposed MAG keep two attributes: \textit{patient} and \textit{interval}. 
The first one identifies the patient; each edge belongs to a single patient, while the nodes might be shared by multiple patients. 
The second one represents the time interval between the origin node and the destination node. 
The time unit of the intervals can vary according to the study case (e.g. hours, days or weeks), but for consistency, all edges of the same model must use the same time unit.

% falar das subdeterminações
Besides being able to comprise the multiple perspectives from which a patient pathway may be assessed, the MAG is also flexible to enable the analysis of different subsets of these perspectives. 
This is possible through the process of \emph{subdetermination}, which consists of converting the MAG into a version based on a subset of its aspects~\cite{Wehmuth2016}. 
For instance, if one decides that the healthcare unit is not relevant for assessing the compliance of a pathway model to a clinical guideline, a subdetermination that disregards the aspect Healthcare Unit would suit the analysis. In this subdetermination, the first edge in Figure~\ref{fig:example-mag}b would become simply \textit{(Antenatal visit, GP, $S_1$, Blood tests, Nurse, $S_2$)} with attributes \textit{patient: Mary} and \textit{interval: 2 days}.

%uma variação interessante seria considerar um mag de cinco aspectos, que além dos 4 apresentados anteriormente, conteria também o diagnóstico atribuído em cada atendimento
Mind that the MAG for modelling patient pathways may have different aspects according to the data available. For example, if the study involved a single hospital, the Healthcare Unit could be suppressed.
It could also be interesting to have the diagnosis as another aspect. The diagnosis field of our dataset, which we present in the next section, has few valid entries. Nonetheless, we also propose the 5-aspect MAG $H^{\prime} = (I,D,O,U,S)$, where the newly-added aspect $D$ is the diagnosis. We will conduct a few experiments with it to test the flexibility of the framework.

% data cleaning
\subsection{Data selection and filtering}
\label{subsec:data}
The data available for this study comprise the public ambulatory medical care provided to the population of São Paulo city, in Southeastern Brazil, from January/2014 to December/2015. There are two files containing the data. In the first one, for each visit, there is information on the medical procedure code (according to the Table of Procedures, Medications, and Orthoses, Prostheses and Special Materials -- SIGTAP), occupation, health care unit, date, patient ID and patient sex. The second file also provides records about the visits including occupation, health care unit, date, patient ID and patient sex, but, instead of containing the medical procedure, it provides the diagnostic code associated with the visit (according to the International Classification of Diseases, Tenth Revision -- ICD-10).

The two case studies we elected for this work are pregnancy pathways and diabetic patient pathways. Besides the fact that we can  satisfactorily assess these cases using only ambulatory data, they pose the main challenges for modelling and mining patient pathways we aim to tackle in this article. 
The recommended pregnancy pathways\footnote{According to the Technical Document of the state of São Paulo available at~\url{www.saude.sp.gov.br/resources/ses/perfil/gestor/atencao-basica/linha-de-cuidado-ses-sp/gestante-e-puerpera/doc_tecnico_quadro_sinteses_e_fluxograma_gestante.pdf}, and the Antenatal Care Protocol (low risk) of the city of São Paulo (available at~\url{www.prefeitura.sp.gov.br/cidade/secretarias/upload/saude/protocolo_PN_baixo_PMSP_13_04_original_modificado_28_4_2023.pdf}). Accessed on 29 aug. 2023~(both documents).} should comprise at least six antenatal care visits spread over the pregnancy period, with increased frequency as the due date approaches. Additionally, according to the guidelines, two obstetric ultrasound scans should take place---one in the first and the other in the second or third pregnancy trimester. Therefore, pregnancy pathways are likely to have a set of interventions repeatedly occurring as time goes on, which is a representation challenge we would like to test our model against.
On the other hand, diabetes is a chronic disease whose evolution depends on various factors, from the genetics of the patients to their lifestyle. Frequently, multidisciplinary teams support the patients, who are also likely to experience multi- and comorbidities. These sources of variability and the demand for understanding not only the medical interventions but also the diagnoses and healthcare professionals involved in the treatment process makes this case study relevant to our goals. 

The following subsections present how we designed and obtained data for each study case.

\subsubsection{Pregnancy study case}
\label{subsec:pregnancy}

In the pregnancy study case, we are interested in analysing the interventions directly related to the pregnancy and not to eventually overlapping pathways, such as allergic rhinitis or tendinitis treatment, which may happen concurrently with the pregnancy period but with no causal relation between them. Therefore, in this section, we present how we selected pregnant patients and how we cleaned their records to focus solely on pregnancy events.

First, we identified patients with at least one diagnostic code or medical procedure exclusively related to pregnancy (see Appendix~\ref{app:cids_sigtaps}) during the six central months of the available period (October/2014 - March/2015), to increase the likelihood of the whole pregnancy being comprised between January/2014 and December/2015. We collected all the visits of these patients, 
totalling \numprint{3245730} records of \numprint{142086} patients. During the data cleaning process, we identified \numprint{429} male patients and \numprint{73} patients who had performed medical procedures and/or received medical diagnoses exclusively applied to male patients.\footnote{According to the table with information about each medical procedure of the SIGTAP coding system -- \url{http://tabela-unificada.datasus.gov.br/tabela-unificada/app/download.jsp}. Accessed on 23 dec. 2022} 
These \numprint{502} patients were removed, resulting in \numprint{3237117} records of \numprint{141584} patients.

Unfortunately, no pair of keys allows us to match the entries of the file containing the medical procedure of the visits to the one containing their diagnosis. When a patient has a single procedure and a single diagnosis on the same date, in the same healthcare unit and provided by the same occupation, it is possible to match the medical procedure to its diagnosis---this 1:1 relation happened in \numprint[\%]{79.13} of the cases. In \numprint[\%]{15.47}, there are two medical procedures and only one diagnosis, so it is not possible to define which procedure the diagnosis refers to. The remaining \numprint[\%]{5.4} cases corresponding to $66$ other combinations. 

\begin{figure}[!ht]
    \centering
    \includegraphics[width=\textwidth]{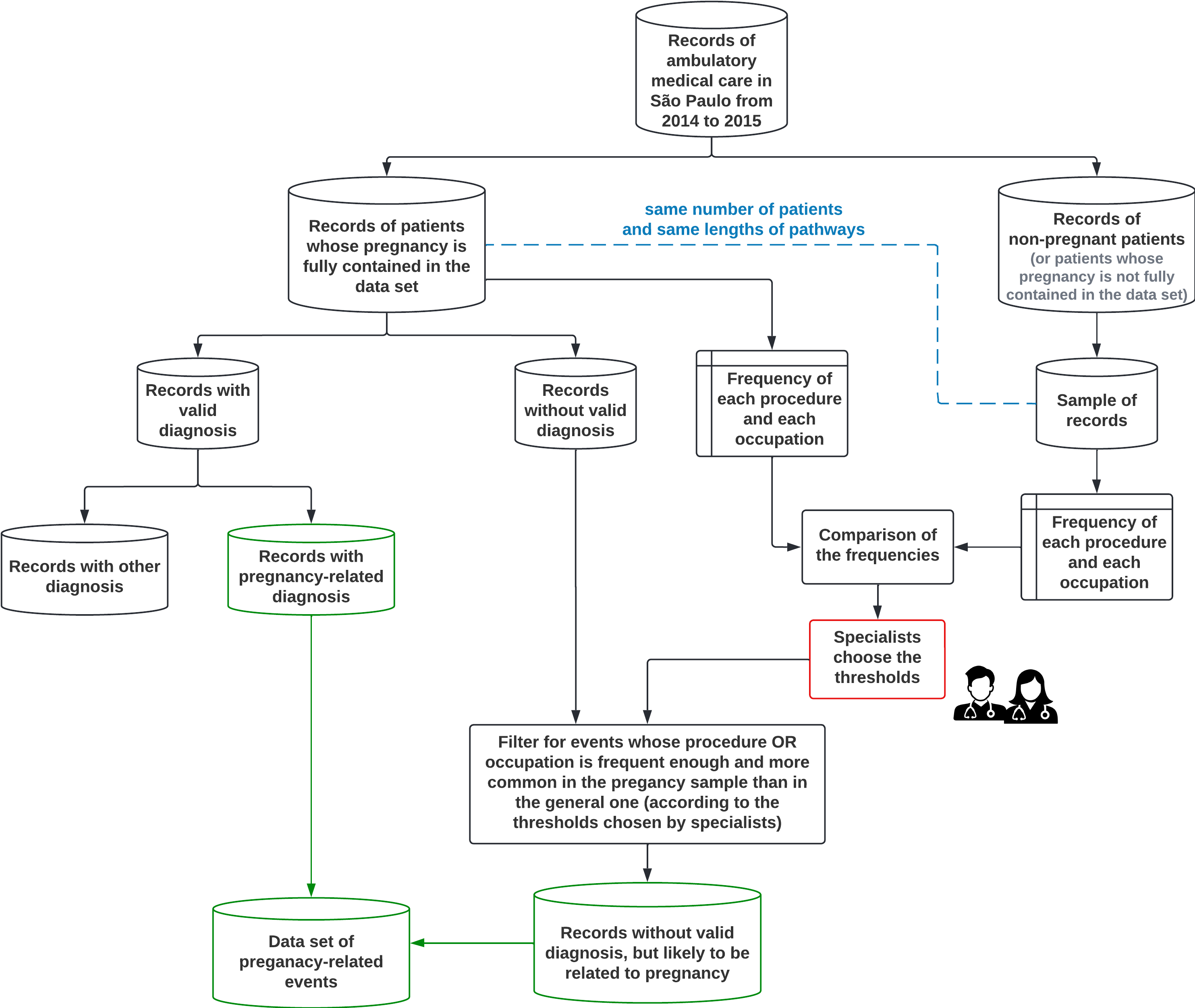}
    \caption{Process to select records related to pregnancy.}
    \label{fig:filtro_gestacao}
\end{figure}

% filter for pregancy events
Although there is a diagnostic field in the dataset, only \numprint[\%]{6.91} of the entries are valid. %it is available for \numprint[\%]{61.5} of the records (\numprint{1990882}), and only \numprint[\%]{11.2} of these are valid entries (\numprint{223664} -- or \numprint[\%]{6.91} of all records). 
As a valid diagnosis is missing for most visits and many medical procedures and occupations are neither specific nor prohibited for pregnancy-related events, it is not a simple task to identify the events directly connected to the pregnancy. \citet{Zaballa2020} faced a similar challenge while looking for medical actions related to the breast cancer treatment pathway. They solved the issue by identifying which medical specialities were more common in the group of patients with at least one record of the diagnoses of interest than in the group of patients without any record of the diagnoses. Then, they kept medical actions whose speciality was among the highlighted ones. We followed a similar approach (Figure~\ref{fig:filtro_gestacao}), but the comparison between the pregnancy and non-pregnancy groups involved both the frequency of medical procedures and the frequency of occupations.

After collecting all records of patients who had a pregnancy during the central months of the available period, we separated the records with valid diagnoses from those with missing or invalid entries. 
From the records with valid diagnoses, we selected the ones whose diagnosis was directly linked to pregnancy. 
To filter the records without a diagnosis, we selected a sample of patients who were not included in the pregnancy cohort to verify the frequency of medical procedures and occupations. 
The sample had the same amount of patients as the pregnancy group and the lengths of pathways were also the same---for example, if there were $10$ pathways of length $5$ in the pregnancy group, then the non-pregnancy sample would also have $10$ pathways of length $5$.\footnote{Four patient pathways from the pregnancy cohort were exceptionally long (\numprint{498}, \numprint{946}, \numprint{1051}, and \numprint{1749} events) and there were no pathways with the same length in the non-pregnancy group. Therefore, we did not consider these four pathways for the frequency comparison analysis.}

\begin{figure}[!ht]
    \centering
    \frame{\includegraphics[width=\textwidth]{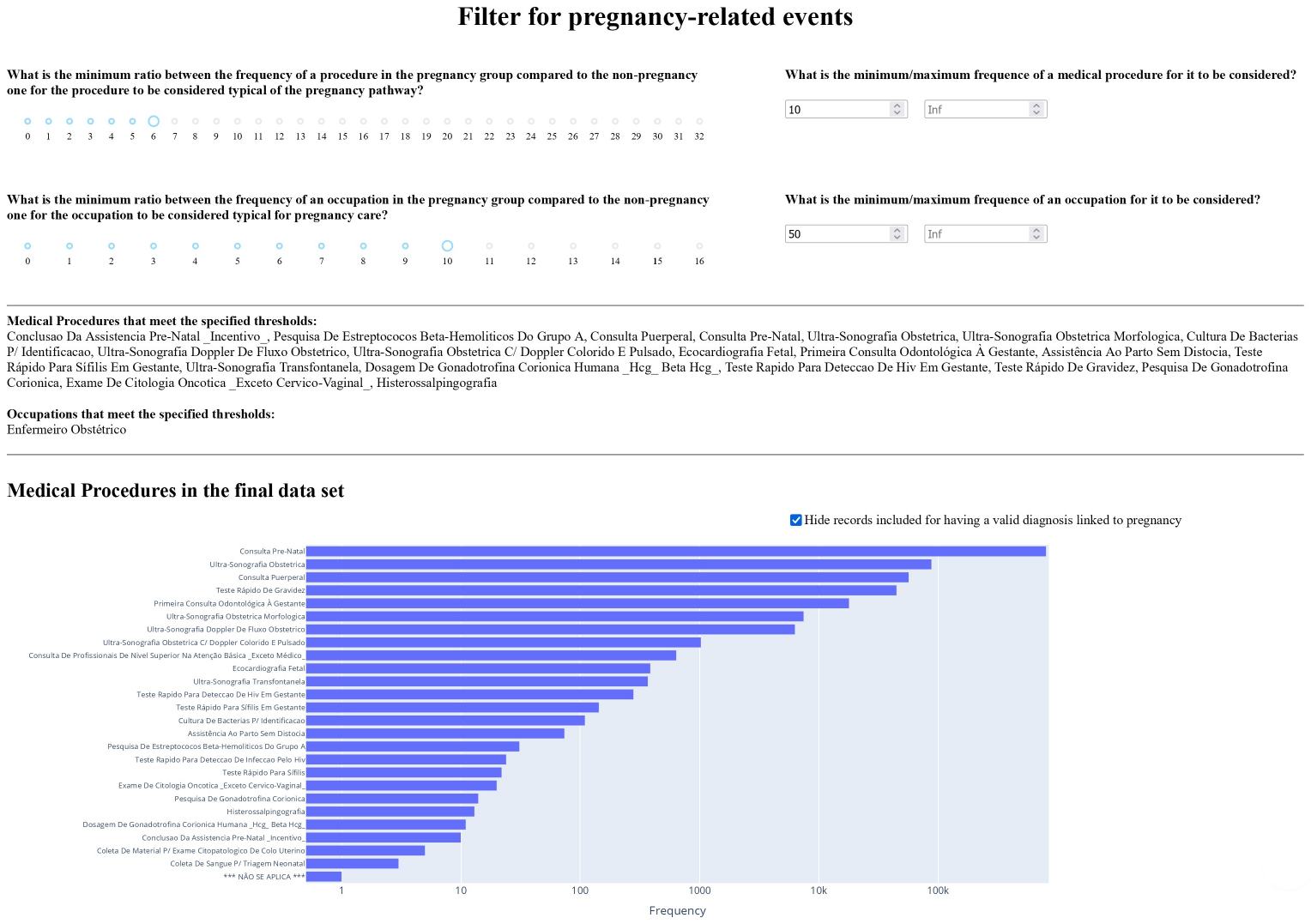}}
    \caption{Interactive tool to support medical specialists in deciding the value of the six proposed thresholds. The medical procedures and occupations listed are in Portuguese, which is the language of the dataset.}
\label{fig:interface_filter_pregnancy}
\end{figure}

We counted the number of occurrences of each medical procedure and each occupation in the pregnancy group and in the non-pregnancy sample. We established six thresholds:
\begin{itemize}
    \item $\theta_{p}$: minimum value of the ratio between the frequency of a medical procedure in the pregnancy group and its frequency in the non-pregnancy group for it to be considered typical of the pregnancy pathway;
    \item $min_{p}$: minimum frequency of a medical procedure in the pregnancy group for it to be included in the list of typical pregnancy procedures;
    \item $max_{p}$: maximum frequency of a medical procedure in the pregnancy group for it to be included in the list of typical pregnancy procedures;
    \item $\theta_{o}$: minimum value of the ratio between the frequency of an occupation in the pregnancy group and its frequency in the non-pregnancy group for it to be considered typical of the pregnancy pathway;
    \item $min_{o}$: minimum frequency of an occupation in the pregnancy group for it to be included in the list of typical  occupations that provide pregnancy care;
    \item $max_{o}$: maximum frequency of an occupation in the pregnancy group for it to be included in the list of typical occupations that provide pregnancy care.
\end{itemize}
A record without a valid diagnosis will take part in the final dataset of pregnancy-related events if its medical procedure satisfies the thresholds $\theta_{p}, min_p, max_p$ or if its occupation satisfies the thresholds $\theta_{o}, min_o, max_o$.
It is a task for the medical specialists to decide the best value for the six thresholds. 
We developed a tool to help them make this decision (Figure~\ref{fig:interface_filter_pregnancy}). 
The interface allows the specialists to vary the values of the thresholds to see which medical procedures and occupations would satisfy them. 
Moreover, it also shows the medical procedures of the events that would pass the filter. 

After filtering the records without diagnosis according to the thresholds, we concatenate them with the records with a valid diagnosis related to pregnancy to obtain the final filtered dataset. 

\subsubsection{Diabetes study case}
\label{subsec:diabetes}

Differently from the pregnancy study case, in the diabetes one, we are interested not only in interventions strictly related to diabetes but also in its comorbidities, such as hypertension. Thus, we do not attempt to filter the pathways, but instead, we keep all records of diabetic patients, independent of which diagnosis they refer to. 
However, many interventions and several medical specialists in the diabetes dataset are not specific to any disease, so, for this case study, we build the same 4-aspect MAG used in the pregnancy case study and also a 5-aspect MAG that includes the diagnosis aspect. In this section, we present how we selected the diabetic patients and built the MAG models.

First, we collected the identifiers of patients with at least one diagnostic code exclusively related to diabetes (see Appendix~\ref{app:cids_diabetes}). 
We extracted all records of these patients in the file containing the medical procedures and in the file containing the diagnoses.

Using the data extracted from the file with medical procedures, we built the 4-aspect MAG proposed in Section~\ref{subsec:mag_model}. 
There are \numprint{71726} patients in the cohort -- \numprint[\%]{60.11} of them  are female, \numprint[\%]{33.07} are male and the sex field is empty for the remaining \numprint[\%]{6.82}. 
There are \numprint{1837518} records associated with diabetic patients, and \numprint[\%]{65.51} of them refer to female patients, \numprint[\%]{30.13} to male patients and \numprint[\%]{4.36} to the patients whose sex information is missing.

To study the comorbidities of the patients, we attempted to merge the records with medical procedures with those containing the diagnoses. %we also built a 5-aspect MAG of diabetic patients, that differs from the former by having the Diagnosis aspect. 
As mentioned in the previous subsection, it is not possible to unmistakably match the entries of the file containing the medical procedures to those of the file containing the diagnoses. However, when we counted the number of medical procedures and diagnoses for the same patient identifier, in the same data, healthcare unit and professional occupation, we found that in \numprint[\%]{83.97} of the cases, there is a single medical procedure and a single diagnosis, in \numprint[\%]{9.01} of the cases there are two medical procedures to a single diagnosis, and in \numprint[\%]{2.00} of the cases, there are three medical procedures to a single diagnosis. Overall, in \numprint[\%]{96.60} cases, there is one or more medical procedures and only one diagnosis. We selected the data related to these cases and merged the records, confirming that none of the merged pairs of medical procedure and diagnosis were incompatible\footnote{According to the table with information about each medical procedure of the SIGTAP coding system -- \url{http://tabela-unificada.datasus.gov.br/tabela-unificada/app/download.jsp}. Accessed on 23 dec. 2022}. 
From the \numprint{1772072} resulting records, \numprint{1269512} (\numprint[\%]{71.64}) had an invalid diagnosis. Therefore, the dataset containing valid entries of both medical procedure and diagnosis has \numprint{502560} records. 
They are concerned with \numprint{69171} patients (\numprint{41765} female, \numprint{22794} male and \numprint{4612} without sex information). Although the number of records significantly dropped after filtering for valid diagnosis entries, we decided to use these data to build a 5-aspect MAG with aspects Intervention, Diagnosis, Occupation, Healthcare Unit and Sequence.
Our goal is to evaluate whether we could find groups of patients according to the multi- and comorbidities and also to highlight the flexibility of our model and mining method for study cases with different features available, such as the diagnosis.

% patient pathways similarity
\subsection{Patient pathway similarity}
\label{subsec:similarity}

In this work, we employ clustering techniques on the patient pathways extracted from the built MAG to mitigate the effect of pathway variability and also to identify relevant pathway patterns. 
To employ these techniques, we must decide on a comparison criterion.

In related studies, \citeauthor{Zaballa2020} and \citeauthor{Kovalchuk2018} used the Levenshtein~\cite{levenshtein1966binary} distance to compare patient pathways. This metric accounts the number of insertions, deletions and substitutions of events to make two patient pathways equal. 
\citeauthor{Zhang2015} used another classic distance metric --- the Longest Common Subsequence~(LCS)~\cite{hirschberg1975linear} --- in their work. \citeauthor{Aspland2021} reviewed $8$ string distance metrics, including Levenshtein and LCS, and proposed a modification of the Needleman-Wunsch algorithm~\cite{Needleman} to compare lung cancer patient pathways. However, none of these metrics contemplate the time of occurrence of the events in each pathway.

As an alternative, \citeauthor{Defossez2014} used the Levenshtein algorithm to evaluate the dissimilarity between breast cancer patient pathways repeating each event of the pathway, which represented the patient state, according to its duration. This strategy indirectly brought time information into the comparison, however, the repetition of characters increases the length of the strings, which also increases the time necessary to compute the distance.

% outra abordagem não relacionada a comparação de sequências usada é a de Rebuge ()Markov) mas que também não levava o tempo em conta

\citeauthor{su2020survey} reviewed trajectory distance measures. They defined trajectory as the ``sequence of time-stamped point records describing the motion history of any kind of moving objects''. Although this does not match the definition of patient pathway, measures that treat trajectories as sequences of discrete observations could also be useful in the patient pathway domain. Among these, the authors identified several measures which they classified as ``sequence-only'' because they contemplate the order of the trajectory points, but not their timestamps. %This category includes the aforementioned Levenshtein and LCS algorithms.
They identified two ``spatial-temporal'' discrete distance measures, which focus both on the order and the timestamp of the trajectory points. One of them measures dissimilarities between sequences of points and timestamps separately, which would not suit patient pathways, but the other deals with them concomitantly. The latter is an extension of the LCS proposed by \citeauthor{vlachos2002discovering}. The algorithm inspect two trajectories to find pairs of points whose coordinates are close enough to each other according to a given threshold and whose timestamps are also close enough according to another threshold. This algorithm could be adapted to the comparison of patient pathways by defining a function that measures how similar two activity tuples are instead of comparing coordinates. Nevertheless, the timestamp comparison is only coherent if the pathways have the same starting point. This constraint forbids us from using the method because we do not have information of when a patient's condition began. In the pregnancy case, for instance, we cannot tell in which pregnancy week the first antenatal care visit of a patient happened, because we do not have neither this information nor the child-birth date.

Therefore, we propose a dissimilarity algorithm based on~\cite{levenshtein1966binary} and \cite{vlachos2002discovering}. Our goal is to compare the activity tuples from both pathways to find those which are similar enough to be considered equivalent. If we find such a compatible pair and consider that the tuples have the same ``meaning'' for both patients, we call it an \emph{alignment} between the pathways. 
During this process, we also take into account the time interval between the activity tuples. 
For example, suppose we align a pair of activity tuples and identify that the subsequent pair is also compatible; however, in one of the pathways, the time interval between the activity tuple that has been aligned and the next one is significantly larger than in the other pathway. 
We should then consider whether this difference in the elapsed time is acceptable. 
If we cannot align an activity tuple from a pathway to any tuple of the other pathway, we increase the value of the distance between the pathways as a penalty. 
Moreover, when an alignment is done but either the activity tuples are not identical or the difference in the time between the last alignment is not zero, we increase the distance between the pathways by an alignment cost that must not be greater than the penalty value.
Mind that it is possible to align a pair of activity tuples after skipping one or more unaligned tuples in one or both pathways. In this case, we should compare the sum of the total accumulated time since the previous alignment instead of simply looking at the last interval.

% To employ these techniques, we establish a comparison criteria, taking into account: 
% \begin{inparaenum}[(i)]
% \item the order of the activity tuples in the patient pathways;
% % \item the similarity between the activity tuples of the pathways being compared---in other words, we want to measure how different two activity tuples are instead of simply saying whether they are equal, i.e. common to both pathways or not;
% \item the time elapsed between the activity tuples of the pathway;
% \item the fact that the pathways being compared might need to be aligned for a proper comparison.
% \end{inparaenum}
% In the related literature, there are several methods for comparing sequences of events~\cite{Egho2015,hirschberg1975linear,levenshtein1966binary,wang2007acs}, but, as far as we are aware, there is no published method that meets all the listed prerequisites. 
% Thus, we developed a dissimilarity measurement that suits our pathway model.

% \subsubsection{Proposed similarity measure}

%Let a patient pathway be expressed by two sequences $S$ and $T$, where $s_{i} \in S$ for $i \leq n$ is the i-th activity tuple in the patient pathway and $t_{i} \in T$ for $i < n$ is the time elapsed between $s_{i}$ and $s_{i+1}$. 
% if the length of the activity sequence ($|S|$) is $n$, the length of the time interval sequence ($|T|$) is $n-1$.  

%To calculate the distance between the pathways $P = \{A,T\}$ and $P^{\prime}=\{A^{\prime},T^{\prime}\}$, we 

Before presenting the algorithm for calculating the dissimilarity between two pathways, we  introduce the following definitions. 
Let $P =\{A,T\}$ be a patient pathway, where $A = (a_{1}, a_{2}, ... , a_{m})$ is the sequence of activity tuples of the pathway and $T = (t_{1}, t_{2}, ... , t_{m-1})$ is the sequence of time intervals---extracted from the corresponding attribute of the edges---such that $t_{i} \in T$ is the time elapsed between $a_{i} \in A$ and $a_{i+1} \in A$.
Let $X = (x_1,x_2,...,x_m)$ be a sequence of $m$ elements, either activity tuples or time intervals. 
The notation $Head(X)$ refers to the first element of the sequence $X$, i.e. $x_1$. 
The notation $Rest(X)$ refers to the subsequence of $X$ containing all its elements except the first one, i.e. $Rest(X) = (x_2,x_3,...,x_m)$; the $Rest(\cdot)$ of the empty sequence is the empty sequence. 
Let $\mathbb{A}$ be the set of all activity tuples; the function $Dist_{A}:\mathbb{A}\times\mathbb{A} \to \mathbb{R}^{+}$ measures how different two activity tuples are. 
It can be defined differently according to the study case, provided that $Dist_{A}(x,y) = Dist_{A}(y,x) \quad \forall x,y \in \mathbb{A}$. 
We define $\delta \in \mathbb{R}^{+}$ as the maximum distance between two activity tuples for them to be considered similar enough to be aligned in the sequences.  
We define $\varepsilon \in \mathbb{R}^{+}$ as the maximum difference between the total time elapsed since the last alignment in the first and second pathways for two events to be aligned. 
When the intervals are not equal, the alignment is possible, but it is associated with a cost ($Dist_{T}$). A suitable way of defining $Dist_{T}$ is dividing the absolute value of the time difference by $\varepsilon$, so that $Dist_{T} = 0$ if, and only if, the intervals are equal, and assumes a maximum value of $1$ if, and only if, the difference between the intervals is the maximum one allowed ($\varepsilon$).
We define $\omega_{A} \in [0,1]$ as the weight that balances $Dist_{A}$ and $Dist_{T}$ for obtaining the total alignment cost of two elements. 
When an alignment is not done, a penalty is returned instead of the alignment cost; its value must be such that it will always be greater or equal to the maximum alignment cost allowed (according to $\delta$, $\varepsilon$ and $\omega_{A}$).

The dissimilarity between the pathways of two patients ($P = \{A,T\}$ and $P' = \{A^{\prime},T^{\prime}\}$) is calculated with the function \textsc{Dissimilrity} presented in Algorithm \ref{alg:distance}. %When the function is called, it returns the cost to align the $Head(\cdot)$ of both pathways (or the penalty value, if they cannot be aligned) plus the minimum distance between the $Rest(\cdot)$ of the pathways.

\begin{algorithm}
\caption{Evaluation of the dissimilarity between two patient pathways}\label{alg:distance}
\begin{algorithmic}[1]\onehalfspacing
\State $t^{}_{ac}$ $\gets$ Null
\State $t^{\prime}_{ac}$ $\gets$ Null
\Function{dissimilarity}{$A$, $A^{\prime}$, $T$, $T^{\prime}$, $t^{}_{ac}$ , $t^{\prime}_{ac}$ }
    \If{$A = \emptyset$ \textbf{and} $A' = \emptyset$}
        \State \Return 0
    \ElsIf{$A = \emptyset$ \textbf{or} $A' = \emptyset$}
        \State \Return penalty + \Call{dissimilarity}{$Rest(A)$,  $Rest(A^{\prime})$, $Rest(T)$, $Rest(T^{\prime})$, Null, Null}
    \ElsIf{$t^{}_{ac}$ = $t^{\prime}_{ac}$ = Null}
        \If {$Dist_{A}(Head(A), Head(A^{\prime})) > \delta$}
            \State \Return \textsc{min}$\big($ \newline
            \hspace*{1em} penalty $+$ \Call{dissimilarity}{$Rest(A)$, $A^{\prime}$, $Rest(T)$, $T^{\prime}$, Null, Null}, \newline
            \hspace*{1em} penalty $+$ \Call{dissimilarity}{$A$, $Rest(A^{\prime})$, $T$, $Rest(T^{\prime})$, Null, Null}\newline
            $\big)$
        \Else
            \State \Return \textsc{min}$\big($ \newline
            \hspace*{1em} $\omega_{A} * Dist_{A}\big(Head(A),Head(A^{\prime})\big) + $\Call{dissimilarity}{$Rest(A)$, $Rest(A^{\prime})$, $Rest(T)$, $Rest(T^{\prime})$, $Head(T)$,  $Head(T^{\prime})$} \newline
            \hspace*{1em} penalty $+$ \Call{dissimilarity}{$Rest(A)$, $A^{\prime}$, $Rest(T)$, $T^{\prime}$, Null, Null}, \newline
            \hspace*{1em} penalty $+$ \Call{dissimilarity}{$A$, $Rest(A^{\prime})$, $T$, $Rest(T^{\prime})$, Null, Null}\newline
            $\big)$
        \EndIf 
    \Else
        \If{$Dist_{A}(Head(A), Head(A^{\prime})) > \delta$ \textbf{or} $|t^{}_{ac} - t^{\prime}_{ac}|>\varepsilon$}
            \State \Return \textsc{min}$\big($ \newline
            \hspace*{1em} penalty $+$ \Call{dissimilarity}{$Rest(A)$, $A^{\prime}$, $Rest(T)$, $T^{\prime}$, $(t^{}_{ac}+Head(T))$, $t^{\prime}_{ac}$}, \newline
            \hspace*{1em} penalty $+$ \Call{dissimilarity}{$A$, $Rest(A^{\prime})$, $T$, $Rest(T^{\prime})$, $t^{}_{ac}$, $(t^{\prime}_{ac}+Head(T^{\prime}))$}\newline
            $\big)$
        \Else
            \State \Return \textsc{min}$\big($ \newline
            \hspace*{1em} $\omega_{A} * Dist_{A}\big(Head(A),Head(A^{\prime})\big)+(1-\omega_{A})*Dist_{T}(t^{}_{ac},t^{\prime}_{ac})\; + $\newline \hspace*{5em}$+$ \Call{dissimilarity}{$Rest(A)$, $Rest(A^{\prime})$, $Rest(T)$, $Rest(T^{\prime})$, $Head(T)$,  $Head(T^{\prime})$} \newline
            \hspace*{1em} penalty $+$ \Call{dissimilarity}{$Rest(A)$, $A^{\prime}$, $Rest(T)$, $T^{\prime}$, $(t^{}_{ac}+Head(T))$, $t^{\prime}_{ac}$}, \newline
            \hspace*{1em} penalty $+$ \Call{dissimilarity}{$A$, $Rest(A^{\prime})$, $T$, $Rest(T^{\prime})$, $t^{}_{ac}$, $(t^{\prime}_{ac}+Head(T^{\prime}))$}\newline
        \EndIf
    \EndIf 
\EndFunction
\end{algorithmic}
\end{algorithm}

If both sequences of activity tuples are empty, there is no pair of events to try to align and there is no need to apply a penalty; thus, the function returns a distance of $0$. This is the base case, as shown in lines 4-5.
On the other hand, if only one of the sequences of activity tuples is empty (i.e., lines 6-7), a penalty is applied and the distance between the $Rest(\cdot)$ of the sequences from both pathways is called. Consequently,  from the moment this if-statement becomes true until both sequences are empty (base case), the number of times the penalty is applied is equal to the length of the non-empty sequence.

Lines 8-12 deal with the cases where both pathways are not empty and there has not been any alignment between them (the variables $t^{}_{ac}$ and $t^{\prime}_{ac}$, which store the time elapsed since the last alignment, are both null). The algorithm evaluates if it is possible to align the first element of $A$ with the first element of $A^{\prime}$---i.e. if $Dist_{A}(Head(A),Head(A^{\prime})) \leq \delta$. 
If this condition is not satisfied (lines 9-10), the algorithm applies a penalty and chooses between skipping the first element of $A$ (calling $Rest(A)$ and $Rest(T)$) while keeping $A^{\prime}$ as it is, or the opposite (skipping the first element of $A^{\prime}$ while keeping $A$ as it is). 
Otherwise (lines 11-12), if it is possible to align $Head(A)$ with $Head(A^{\prime})$, the algorithm evaluates whether to align or to skip an element in one of the sequences for fostering a better alignment between subsequent elements.

After the first alignment is made, besides calculating $Dist_{A}$ it is necessary to consider the time elapsed since the last alignment in both pathways (lines 14-19). 
Lines 15-16 show that if the activity tuples are not similar enough ($Dist_{A}(Head(A),Head(A^{\prime}))>\delta$) or the difference between the time elapsed since the last alignment is too large ($|t^{}_{ac} - t^{\prime}_{ac}|>\varepsilon$), it is not possible to align $Head(A)$ and $Head(A^{\prime})$ and, similarly to the case with no previous alignment, the algorithm must choose between skipping an element in $P$ or $P^{\prime}$. The difference is that besides calling $Rest(A)$ and $Rest(T)$ (or $Rest(A^{\prime})$ and $Rest(T^{\prime})$), the value of $t^{}_{ac}$ (or $t^{\prime}_{ac}$) must increase by adding the value $Head(T)$ (or $Head(T^{\prime})$), which corresponds to the time interval between the event being skipped and the following one. 
On the other hand (lines 17-18), if the activities are compatible for alignment and the difference of time elapsed since the last alignment is within the specified limit, the algorithm must verify whether the alignment or skipping $Head(A)$ (or $Head(A^{\prime})$) yields a smaller value. 
To obtain the alignment cost, the algorithm calculates the distance between the activity tuples ($Dist_{A}(Head(A),Head(A^{\prime}))$) and the distance based on the difference between the time elapsed since the last alignment ($Dist_{T}(t^{}_{ac},t^{\prime}_{ac})$), and weighs these values using the parameter $\omega_{A}$. 
It then calls the distance function for the $Rest(\cdot)$ of both pathways, but resets $t^{}_{ac}$ and $t^{\prime}_{ac}$ so that they store the time intervals between the events it has just aligned and the subsequent ones in both pathways ($Head(T)$ and $Head(T^{\prime})$, respectively).

We used the proposed measurement %computed by Algorithm \ref{alg:distance} 
to calculate pairwise distances between \numprint{2000} patient pathways whose length was less than or equal to the third quartile of pathway length. We defined two functions to calculate the distance between the activity tuples of the pathways ($Dist_A$)---one for the 4-aspect MAG and the other, for the 5-aspect MAG.
First, given two activity tuples (nodes) of the 4-aspect MAG, $v_1$ and $v_2$, expressed as $(v^i_1,v^o_1,v^u_1,v^s_1)$ and $(v^i_2,v^o_2,v^u_2,v^s_2)$ respectively, where $v^i_1$ and $v^i_2$ are elements of the aspect Intervention,  $v^o_1$ and $v^o_2$ are elements of the aspect Occupation,  $v^u_1$ and $v^u_2$ are elements of the Healthcare Unit aspect, and  $v^s_1$ and $v^s_2$ are elements of the aspect Sequence, we defined $Dist_{A}$ between them with Equation~\ref{eq:distA_1}.

\begin{equation}
    \label{eq:distA_1}
    Dist_{A}(v_1,v_2) = 
    \begin{cases}
        0 &  \text{ if } v^i_1 = v^i_2 \text{ and } v^o_1 = v^o_2\\
        0.3 & \text{ if }  v^i_1 = v^i_2 \text{ and } v^o_1 \neq v^o_2\\
        0.7 & \text{ if } v^i_1 \neq v^i_2 \text{ and } v^o_1 = v^o_2 \\
        1 & \text{ if } v^i_1 \neq v^i_2 \text{ and } v^o_1 \neq v^o_2\\
    \end{cases}.
\end{equation}

As regards the 5-aspect MAG, given two activity tuples, $v_1$ and $v_2$, expressed as $(v^i_1,v^d_1,v^o_1,v^u_1,v^s_1)$ and $(v^i_2,v^d_2,v^o_2,v^u_2,v^s_2)$ respectively, where $v^i_1$ and $v^i_2$ are elements of the aspect Intervention, $v^d_1$ and $v^d_2$ are elements of the aspect Diagnosis,  $v^o_1$ and $v^o_2$ are elements of the aspect Occupation,  $v^u_1$ and $v^u_2$ are elements of the Healthcare Unit aspect, and  $v^s_1$ and $v^s_2$ are elements of the aspect Sequence, we defined $Dist_{A}$ between them with Equation~\ref{eq:distA_2}.

\begin{equation}
    \label{eq:distA_2}
    Dist_{A}(v_1,v_2) = 0.7 *
    \begin{cases}
        0 & \text{ if all 4 digits of } v^d_1  \text{ match those of } v^d_2 \\
        0.3 & \text{ if the first 3 digits of } v^d_1  \text{ match those of } v^d_2 \\
        1 & \text{ otherwise}\\
    \end{cases} + 0.3 *
    \begin{cases}
        0 & \text{ if } v^i_1 = v^i_2 \\
        1 & \text{ if } v^i_1 \neq v^i_2 \\
    \end{cases}
\end{equation}

The comparison of the diagnosis component of the activity tuples involves the number of digits because our dataset provides the diagnoses according to the ICD-10 system, which hierarchically organises its codes. If the four digits are equal, the diagnoses are equal; if only the first three digits are equal, the diagnoses are not equal but belong to the same disease. Although some diseases are fully characterised by a 3-digit code, without 4-digit versions, we do not treat these cases separately.  

%We considered the scenarios in Table \ref{tab:scenarios_dissimilarity}. 
%We evaluated each scenario with/without filtering the pathways for events exclusively related to pregnancy. 

% \begin{table}[!hbt]
%     \centering
%     \begin{tabularx}{\textwidth}{c|X|c|c}
%         \hline
%         \textbf{Scenario} & \textbf{Criterion for pairing activity tuples}  & $\mathbf{\delta}$ &  $\mathbf{\varepsilon}$  \\ \hline\hline
%         I & If the procedures and occupations are equal, $Dist_{A} = 0$; otherwise, $Dist_{A} = 1$ & $0.99$ & $20$ days \\ \hline
%         II & \noindent\parbox[c][50pt]{\hsize}{  If  the procedures and the occupations are equal, $Dist_{A} = 0$; if the procedures are equal and the occupations different, $Dist_A = 0.3$;
%         if the occupations are equal and the procedures different, $Dist_A = 0.7$; otherwise, $Dist_A = 1$} & $0.5$ & $20$ days  \\ \hline
%         III & \noindent\parbox[c][25pt]{\hsize}{If  the procedures, occupations and healthcare units are equal, $Dist_{A} = 0$; otherwise, $Dist_A = 1$} & $0.99$ & $20$ days \\ \hline
%     \end{tabularx}
%     \caption{Tested configuration of parameters to calculate the distances between the pathways.}
%     \label{tab:scenarios_dissimilarity}
% \end{table}

% patient pathway clustering
\subsection{Patient pathway clustering}
\label{subsec:clustering}

\begin{figure}[!ht]
    \centering
    \includegraphics[width=0.85\textwidth]{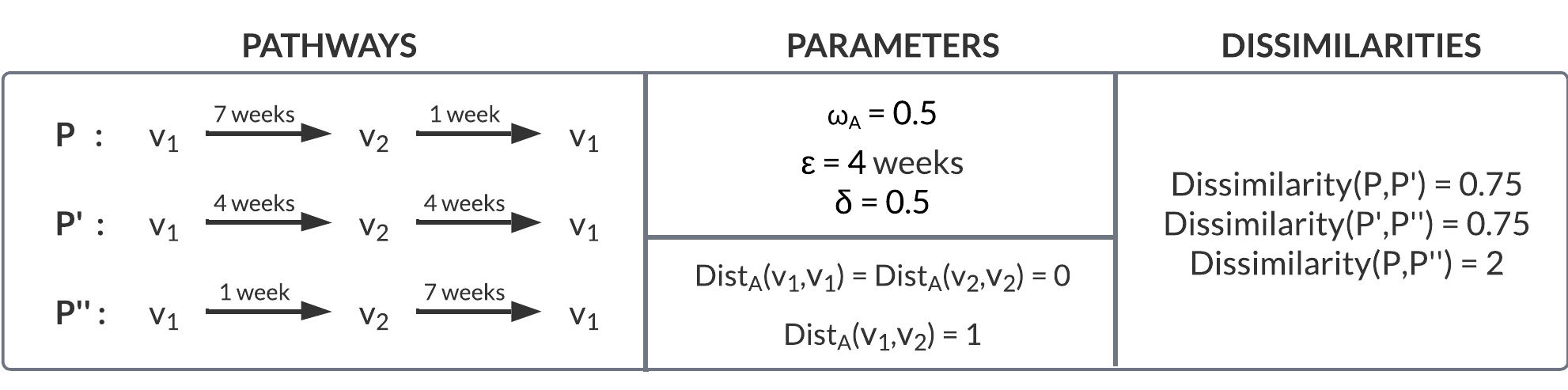}
    \caption{Example of three pathways whose dissimilarities do not satisfy the triangle inequality (with the given parameters).}
\label{fig:example_triangle_inequality}
\end{figure}

The calculated dissimilarities can now feed a clustering method to find groups of similar pathways.
The method must be compatible with a pre-computed dissimilarity matrix. Some well-established examples are DBSCAN, k-medoids and hierarchical clustering (single, complete or average linkage).
However, the dissimilarity measurement proposed in the last subsection does not satisfy the triangle inequality as a result of the time interval restriction to perform the alignments. 
In other words, it is possible to have two dissimilar pathways which are both similar to a third one. 
Figure~\ref{fig:example_triangle_inequality} displays an example where this happens. 
Suppose three diabetic patients had a medical visit with an endocrinologist~($v_1$); all of them were recommended to seek a dietitian's advice~($v_2$) before the follow-up with the endocrinologist. The first patient ($P$) took 7 weeks to have an appointment with the dietitian---only a week before the second endocrinologist's visit; the second patient ($P^{\prime}$) sought the dietitian advice 4 weeks after the first endocrinologist's visit---4 weeks before the second one; the third patient visited the dietitian shortly after (1 week) the endocrinologist's recommendation and had 7 weeks to change her diet before returning to the endocrinologist. If we consider that 4 weeks is the maximum difference between time intervals to align events of two pathways, we would be able to align all visits to the endocrinologist between the pathways; the dietitian visits of patients $P$ and $P^{\prime\prime}$ have a difference of $6$ weeks since the first alignment and cannot be matched; however, the dietitian visits of patient $P^{\prime}$ can be aligned with the one of pathway $P$ and with the one of pathway $P^{\prime\prime}$. As a result, the dissimilarity between the pathways $P^{\prime}$ and $P$ and the dissimilarity between $P^{\prime}$ and $P^{\prime\prime}$ are small when compared to the dissimilarity between $P$ and $P^{\prime\prime}$ -- indeed, the sum of the first two values is smaller than the latter.

Moreover, the pathways of patients with chronic conditions are frequently affected by several comorbidities, so, taking again the example of diabetes, it is feasible to find patients with diabetes and hypertension, patients with diabetes and asthma, and patients with diabetes, asthma and hypertension. In this example, the patient pathways of the third group would be similar to those in the first two. 
With the unfulfilled triangle inequality of the dissimilarity measurement and the comorbidity likelihood of chronic conditions working together, it becomes harder to assign a patient to only one group, where all the pathways are very similar to it while guaranteeing that the pathway is significantly different from those in the other groups---which is the basic assumption of traditional hard clustering methods.  
Alternatively, it would be more suitable to have a flexible classification that assumes a patient has a probability of belonging to a certain group or, at least, that acknowledges that their pathway may share characteristics of several identified patterns.
This points to soft clustering approaches~\cite{ref_fuzzy_clustering}, also called fuzzy clustering, or to the task of detecting overlapping communities of nodes in a network~\cite{survey_overlappig_communities}.

The Order Statistics Local Optimization Method (OSLOM) algorithm~\cite{oslom} is an example of method that detects overlapping communities in graphs. It looks for subgraphs whose nodes are more strongly connected to each other than one would expect to find in a random network with the same degree and strength (if the network is weighted). 
Starting from a randomly selected node, the method picks some of its neighbours to define an initial community. Then, for each node outside the community but linked to it by at least one edge, the algorithm verifies how likely it would be to find a node at least as strongly connected to the community as the evaluated one is in a random network, where no communities are expected to arise. The nodes with the smallest probability are added to the community, and after that, the method checks if the nodes in the updated community are still significant to it; those that are not are removed. These two steps of assessing neighbours and members of the community are repeated until the community remains unchanged. 
The algorithm repeats the whole procedure starting from as many different nodes as necessary to reach similar communities repeatedly. Then, it evaluates the clusters to find a set of significant minimal ones. As the process of detecting communities starting from a node is independent of the detection starting from any other, a node may end up taking part in more than just one community. Moreover, the method does not require every node to be assessed to at least one community, highlighting ``homeless'' vertices. It is also able to detect higher-level communities among the found clusters~\cite{oslom}.

OSLOM suits patient pathway clustering as proposed here, because it supports weighted edges, so the complement of the precomputed dissimilarity measure works as the adjacency matrix of a graph whose overlapping communities of nodes (pathways) we seek. We opted to add only edges with weight greater than zero (distance less than $1$, the maximum value possible) to the graph.

%The calculated distances can now feed a clustering method that is compatible with a pre-computed dissimilarity matrix.
%Examples of well-established methods with this characteristic are DBSCAN, k-medoids and hierarchical clustering.
%However, the dissimilarity measurement proposed in the last subsection does not satisfy the triangle inequality

%There are several options for such a clustering method; we opted for hierarchical clustering.
%For each scenario in Table \ref{tab:scenarios_dissimilarity}, we used an agglomerative algorithm and compared the cophenetic correlation of single, complete and average linkage to choose the option with the highest correlation to build the dendrogram.
%We used the Mojena criterion~\cite{Mojena1977,Milligan1985} with a critical value of \numprint{1.25} to cut the dendrogram and obtain the clusters. 

% centrality filter for simplifying the pathway model
\subsection{Patient pathway model simplification}
\label{subsec:simplification}

Even though clustering patient pathways helps to mitigate their high variability, some clusters might still bear many different behaviours. Thus, we developed a technique for simplifying the MAG and facilitating its interpretation. 
This technique consists of filtering nodes from the pathway model according to a measure of the node relevance. 
We assume that a node is relevant if it is part of the core pathway followed by most patients, or if it corresponds to a worsening branch of the pathway model.
As our model supports multi-perspective analyses, we consider how each of them might contribute to the relevance of the nodes. 
We take advantage of the isomorphism between the MAG and the traditional graph~\cite{Wehmuth2016} to apply classical centrality metrics to specific subdeterminations of the proposed MAG.
The following subsections present which centrality measure we used in each subdetermination, and how we combined them to obtain the relevance of the activity tuples. Some of the metrics depend on predetermined parameters---in these cases, we will discuss their values in Section~\ref{sec:results}.

\subsubsection{Influence of the Healthcare Unit on the node relevance}
First, when it comes to healthcare units, the Brazilian health system is organised as a healthcare network composed by healthcare units with different levels of specialisations.%\atencao{incluir informação sobre as unidades e seus tipos}. 
Thus, in a network whose nodes are healthcare units, it is reasonable to expect that the more specialised a unit is, the larger the number of the units sending patients to it. 
Moreover, highly specialised units will receive patients from a broad range of low-specialised and moderately specialised units; the moderately specialised units will receive patients from several low-specialised units; and the low-specialised units will receive patients from a few other low-specialised units. %and the specialisation level of the units sending patients to it. 
To obtain such a network, we can take the subdetermination of the proposed MAG that keeps only the Healthcare Unit aspect.
%This behaviour could then be captured by a method that measures the importance of a node based on the number and importance of the nodes pointing to it.

In the pregnancy study case, the fact that a patient requires assistance from specialised units is an indicator that something out of the ordinary might be happening in their pathway. 
Therefore, we use the PageRank algorithm to calculate the relevance (as an estimate of the degree of specialisation) of the units according to the number and relevance of the units sending patients to them. 

The PageRank algorithm~\cite{pagerank} measures the importance of a node in a network based on the importance of the nodes pointing to it. 
However, if an important node points to many others, the gain of importance of the successors is smaller than it would be if the important node had fewer outgoing edges. 
PageRank defines the relevance $C_{pgr}(i)$ of the node $i$ as

\begin{equation}
\label{eq:PR_semBeta}
    C_{pgr}(i) = \sum_{j=1}^{N} C_{pgr}(j) \bar{A_{ji}}
\end{equation}

where $N$ is the number of nodes in the network and $\bar{A_{ji}}$ is a modification of the value of the adjacency matrix between nodes $j$ and $i$ as expressed in Equation \ref{eq:adjacencia_grau}. 
Each entry of $\bar{A_{ij}}$ corresponds to the value of the adjacency between node $i$ and $j$ divided by the out-degree of the origin node ($k_i^{out}$) if it is not zero; if the out-degree is zero, $A_{ij}$ is also zero and, to prevent a case of zero divided by zero, $\bar{A_{ij}}$ is defined as zero in this case.
Using $\bar{A_{ij}}$ instead o $A_{ij}$ dilutes the importance of a node between its successors, instead of assigning the full value to each successor. 

\begin{equation}
\label{eq:adjacencia_grau}
    \bar{A}_{ij} = 
    \begin{cases}
        \frac{A_{ij}}{k^{out}_{i}} & \quad \text{if } k^{out}_{i} > 0 \\
        0 & \quad \text{if } k^{out}_{i} = 0
    \end{cases}.
\end{equation}

%where $k^{out}_{i}$ is the out degree of node $i$.

A shortcoming of Equation~\ref{eq:PR_semBeta} is that, if a node $i$ has no incoming edges, its importance is zero ($C_{pgr}(i) = 0$), and each node to which $i$ points will receive no increment in its importance from $i$. As an alternative, each node can receive a ``free'' centrality value coming from  a constant term $\beta_i$, as in Equation~\ref{eq:PageRank}. 

\begin{equation}
\label{eq:PageRank}
    C_{pgr}(i) = \alpha \sum_{j=1}^{N} C_{pgr}(j) \bar{A}_{ji} + \beta_i
\end{equation}

where $\alpha$ is a constant which balances the importance of the first term of the equation when compared to the second one.
The value $\beta_i$ might be the same for every node, or it can be personalised to represent the inherent importance of each node, regardless of the network structure. 

%It is also possible to have nodes without outgoing edges, which are often called ``dangling nodes''. Their existence causes matrix $\mathbf{\bar{A}}$ to have rows whose entries are all zero. All the other rows (non dangling nodes), by definition, sum up to one. Actually, each entry $a_j$ from the $i-th$ row can be interpreted as the probability from transitioning from node $i$ to node $j$. If we convert the rows with only zero entries to the vector $\frac{1}{n}\mathbf{1}$, where \mathbf{1} is the unitary vector of length $n$, then the matrix $\mathbf{\bar{A}}$ becomes a row stochastic matrix.
%As there is a circular dependency between the importance of the nodes, if we guarantee the stochastic matrix is also irreducible, we can solve the expression iteratively, knowing that there will be a unique stationary distribution (the centrality vector).

%We used the NetworkX implementation of the PageRank algorithm \cite{networkx} and set $\alpha = 0.85$, which is a common choice for the parameter~\cite{surveyPagerank,bookNewman}, and $\beta = \frac{1}{N}$ for each node, where $N$ is the number of nodes in the network. % default 1/N

\subsubsection{Influence of the Occupation on the node relevance}
It is common for a lead doctor/medical specialist to assist the patients and, eventually, refer them to other professionals, according to their needs. 
In a network whose nodes are the occupation of the healthcare professionals, the nodes playing the central role in the  pathways are more likely to have several connections (edges) to other occupations, while those who play an auxiliary role would be less connected to the remaining ones. 
As a result, the central role nodes would tend to mediate the connections between the auxiliary nodes. 
In complex network theory, the betweenness centrality measures the importance of a node according to the number of paths between the other nodes in the network that pass through that node.  
Formally, let $g_{uv}$ be the number of geodesic paths (shortest paths) between the nodes $u$ and $v$ in a graph $G = (V,E)$, and let $g^{i}_{uv}$ be the number of those paths that include the node $i$. The betweenness centrality of the node $i$ is defined as the sum of the fractions $\frac{g^{i}_{uv}}{g_{uv}}$ for each pair of nodes $(u,v)$ in the network, as given by Equation~\ref{eq:betweenness}.

\begin{equation}
\label{eq:betweenness}
    C_{bet}(i) = \sum_{u,v \in V} \frac{g^{i}_{uv}}{g_{uv}}   
\end{equation}

Therefore, as an effort to identify events pertaining to the core pathway, we evaluate the betweenness centrality of the subdetermined version of the proposed MAG that keeps only the aspect Occupation, to measure how central an occupation is to the model. % We adopted the NetworkX implementation of the betweenness centrality~\cite{networkx}.

\subsubsection{Influence of the Intervention on the node relevance}
The intervention can also support the identification of the core pathway. 
Instead of finding interventions playing central roles, as in the case of the occupations, the interest lies on the identification of those that appear repeatedly in the patient pathways. 
This repetition increases the trend for the intervention to be close to (occur immediately before or after) a broader range of other interventions. 
In a directed graph where the interventions are nodes and the edges represent interventions that directly followed one another in at least one pathway, the nodes that can reach others in fewer steps are the ones we are interested in. 
The closeness centrality uses a similar logic to measure the importance of a node---the nodes with higher closeness centrality measures are those whose distances to the remaining nodes are generally shorter.
Formally, considering the graph $G = (V,E)$ and the node $i \in V$, let $V_{i} \subset V$ be the set of nodes $j \in V$, such that $j$ can be reached from $i$ and $j \neq i$. Let $N_i$ be the number of nodes in $V_i$. The closeness centrality is given by Equation~\ref{eq:closeness1}.

\begin{equation}
    \label{eq:closeness1}
    C_{clo}(i) = \frac{N_i}{\sum\limits_{j \in V_i}d_{ij}}
\end{equation}

where $d_{ij}$ is the shortest-path distance between nodes $i$ and $j$. Thus, the closeness centrality of node $i$ is given by the inverse of the average minimum distance between node $i$ and the other nodes in the network which can be reached from $i$. This assigns higher values for vertices closer to the remaining vertices of the network.
However, the expression in Equation~\ref{eq:closeness1} might not allow a proper comparison of the centrality measure of nodes in an unconnected graph (or a non-strongly connected digraph), as the value of $N_i$ varies among the nodes. As an alternative, the centrality value of a node $i$ can be multiplied by the fraction of all nodes in the network that can be reached from it~\cite{wasserman_faust_1994}, as given by Equation~\ref{eq:closeness2}.

\begin{equation}
    \label{eq:closeness2}
    C_{clo}(i) = \left(\frac{N_i}{N-1}\right)\frac{N_i}{\sum\limits_{j \in V_i}d_{ij}}
\end{equation}

where $N$ is the number of nodes in $G$, and, thus, $N-1$ is the maximum number of nodes a node could reach.  

We applied this centrality metric to the nodes of the subdetermined version of the proposed MAG to keep only the Intervention aspect, to identify interventions linked to the core pathway. 
% We adopted the NetworkX implementation of the closeness centrality~\cite{networkx}.
%are ``closer'' to the others, either because the procedure appears in different moments of the pathways, thus having several neighbours, or because the procedure is a frequent neighbour of such a node.
% closeness para MAG faria mais sentido, não??

\subsubsection{Resulting relevance of the nodes}
The relevance of the node $V$ expressed as $(v^i,v^o,v^u,v^s)$, where $v^i$ is an element of the aspect Intervention, $v^o$ is an element of the aspect Occupation, $v^u$ is an element of the Healthcare Unit, and $v^s$ is an element of the aspect Sequence, could then be expressed as in Equation \ref{eq:relevancia_inicial}.

\begin{equation}
\label{eq:relevancia_inicial}
    R_0\big((v^i,v^o,v^u,v^s)\big) = w_1*C_{pgr}(v^u) + (1-w_1) * \big(w_2  * C_{clo}(v^i) + (1-w_2)*C_{bet}(v^o)\big)
\end{equation}

where:
\begin{itemize}
    \item $C_{clo}(v^i)$ is the closeness centrality of the intervention $v^i$ in the subdetermined version of the model that keeps only the Intervention aspect; 
    \item $C_{bet}(v^o)$ is the betweenness centrality of the occupation $v^o$ in the subdetermined version of the model that keeps only the Occupation aspect; 
    \item $C_{pgr}(v^u)$ is the PageRank centrality of the healthcare unit $v^u$ in the subdetermined version of the model that keeps only the Healthcare Unit aspect; 
    \item $w_2$ is a real number ($0 \leq w_2 \leq 1$) that weighs the contribution of the closeness of the $v^i$ component of the node and the betweenness of the $v^o$ component, as both of them contribute to the relevance of the node in terms of how likely it is to be part of the typical pathway; and
    \item $w_1$ is a real number ($0 \leq w_1 \leq 1$) that weighs the importance of the combination of centralities regarding the typical pathway ($w_2  * C_{clo}(v^i) + (1-w_2)*C_{bet}(v^o)$) and the PageRank centrality of $v^u$, which highlights the most specialised healthcare units, where patients with medical complications are more likely to receive treatment. %nodes likely to take part in a complication of the pathway.
\end{itemize}

It is worth noticing that Equation \ref{eq:relevancia_inicial} gives the same relevance to activity tuples with the same intervention, occupation and healthcare unit, not taking into account when they happened (aspect Sequence) and in which context (predecessor and successor nodes).
Alternatively, instead of using $R_0$ as the only measure of relevance of the nodes in the pathway model, we use it to feed the constant term $\beta$ of the PageRank algorithm for this algorithm to calculate the influence of the Sequence and the context over the nodes. 
In order to use the PageRank algorithm in the complete (non-subdetermined) MAG, we create artificial edges in the opposite direction of the real ones, so that a node will be important whether it happens before or after an important one---e.g. first action after (or last intervention before) a life-threatening event.
%We used the NetworkX implementation of the PageRank algorithm \cite{networkx}, and experimentally found that a value of $\alpha$ around $0.2-0.3$ led to reasonable results, according to the interpretation of the mined model by a medical specialist. % In fact, we developed a tool for visualising the effects of each parameter on the node relevance and the resulting model for each cluster. --> Results?

% ----------------------------------------------
\section{Results}
\label{sec:results}

We now present the results of modelling, clustering and mining patient pathways of the pregnancy and the diabetic case studies. 
The framework implementation is available in a GitHub repository.~\footnote{Available at \url{https://github.com/caroline-rosa/framework_patient_pathways}.}
Although the original data are in Portuguese, we translated the interventions and occupations of tables and figures of this section into English to facilitate the interpretation. 

\subsection{Data preparation and inspection}

In the pregnancy case study, we filtered the records of the patients to focus on events directly related to their pregnancy. 
The filter parameters were tuned with supervision of coauthor MI, who is a medical doctor. 
The thresholds used to obtain the MAG filtered for pregnancy-related events were $\theta_p = 6$ and $\theta_o = 10$. The minimum frequency for interventions to be selected was $min_p = 10$ and the minimum frequency for occupations to be selected was $min_o = 50$. We did not set a maximum value neither for the intervention frequency nor for the occupation frequency ($max_p = max_o = \infty$). We also built a MAG before applying this filter, for the sake of comparison and also because %the medical specialists who assisted our work pointed out that, 
for some analyses, medical specialists could also be interested in assessing what happened before and after the pregnancy. %This could allow investigating, for instance, whether a woman with diabetes mellitus in pregnancy had the condition before it, or if she remained diabetic after the puerperium.

\begin{figure}[!ht]
    \centering
    \includegraphics[width=0.9\textwidth]{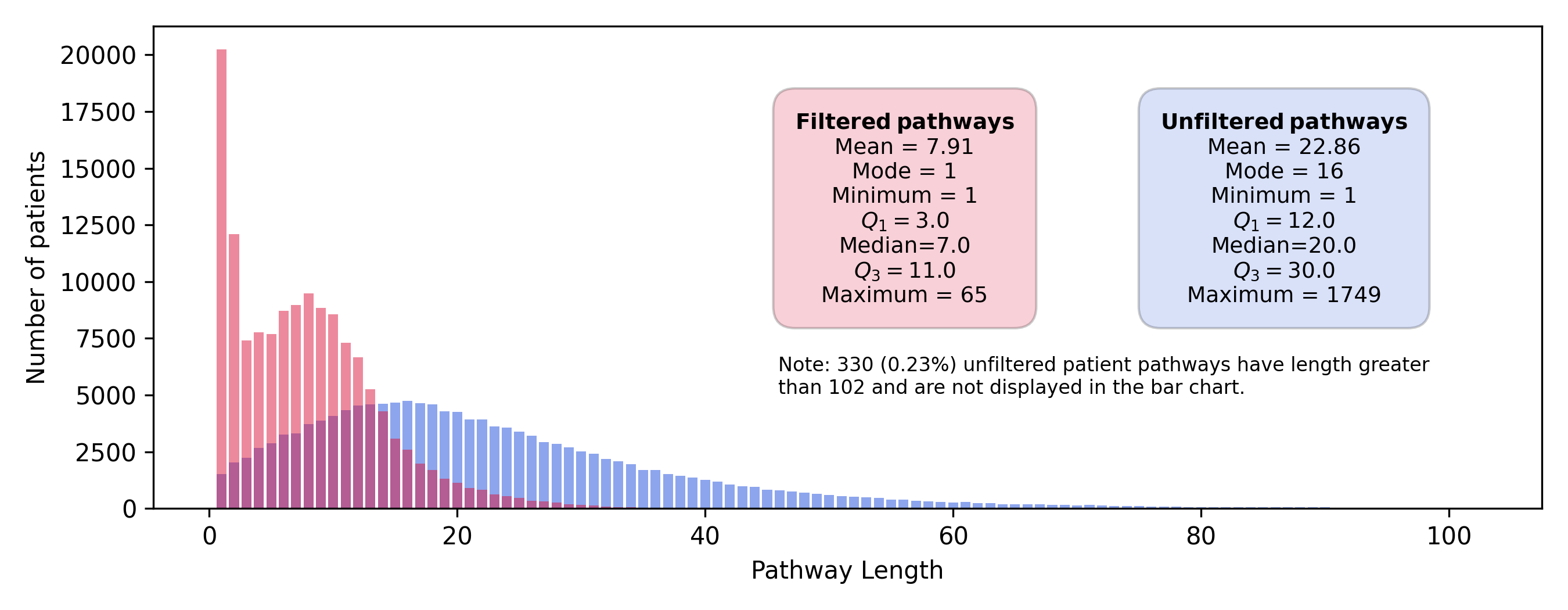}
    \caption{Distribution of patient pathway length with and without filter for pregnancy-related events only.}
    \label{fig:tamanho_jornadas}
\end{figure}

Figure~\ref{fig:tamanho_jornadas} presents the distribution of pathway length (number of activity tuples) before and after filtering for pregnancy-related events.
The set of unfiltered pathways has \numprint{141584} patients, whose median pathway length is $20$ events. The lower quartile ($Q_1$) is $12$ and the upper quartile ($Q_3$) is $30$, leading to an Interquartile Range (IQR) of $18$.
Nevertheless, there are a few pathways ($330$, or \numprint[\%]{0.23}) whose length is significantly higher than the others (greater than $Q_3+4*IQR$), having as many as \numprint{1749} events. While this longest pathway could be the result of a data quality issue, when a clinician evaluated it, it turned out to be a likely legitimate path for a patient who required extensive psycho-social service attention.

After filtering the pathways to focus on pregnancy-related events, the median length of the pathways decreased from $20$ to $7$ because the filtering process left out events not deemed as pregnancy-related. The distribution of patient length also became more condensed and its shape changed, with emphasis on the number of pathways with only $1$ or $2$ events.
The set of filtered pathways has \numprint{140159} patients. The number of patients decreased by \numprint{1425} when compared to the unfiltered group, which means none of the events of these patients were accounted as part of their pregnancy pathway. This happened because \begin{inparaenum}[(i)]
    \item $977$ patients had medical procedures belonging to the list used to select pregnant patients to the case study cohort~(Appendix~\ref{app:cids_sigtaps}), but associated with non-pregnancy diagnoses, thus not selected by the filter;
    \item $439$ had at least one pregnancy diagnostic code used to obtain patients to the case study cohort, but it was not possible to match them to any of the patient’s medical procedures, which were not specific to pregnancy themselves;
    \item $7$ patients matched both previous cases;
    \item $2$ patients entered the case study cohort for having medical procedures linked to pregnancy, but none of them satisfied the filter thresholds.
\end{inparaenum}

\begin{figure}[!ht]
    \centering
    \includegraphics[width=0.9\textwidth]{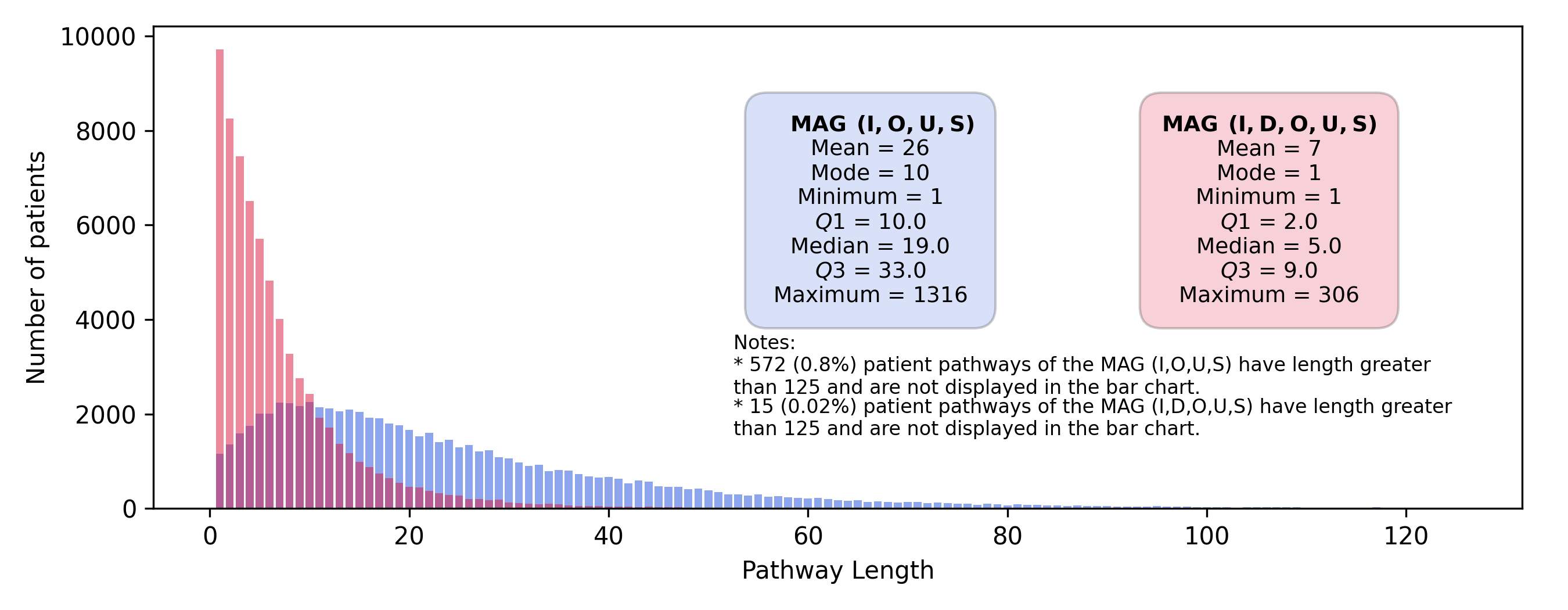}
    \caption{Distribution of patient pathway length of diabetic patients in the 4-aspect MAG and in the 5-aspect MAG.}
    \label{fig:tamanho_jornadas_diabetes}
\end{figure}

In the diabetic study case, we did not filter the records strictly related to the treatment of diabetes, but, as the records with available and valid diagnosis information were limited, the number of pathways in the 5-aspect diabetes MAG and their distribution of length significantly differed from those of the 4-aspect diabetes MAG (see Figure~\ref{fig:tamanho_jornadas_diabetes}).

The patient pathways of the 4-aspect MAG have a median length of 19 events and the pathways of the 5-aspect MAG have a median length of 5 events.

\subsection{Modelling the patient pathways as a MAG}

Information about each MAG is shown in Table~\ref{tab:mag_info}. The overall reduction in the length of the pathways brought by the filter in the pregnancy case study and by the diagnosis availability in the diabetes study case is also reflected in a drop in the number of nodes and edges between the models. 

\begin{table}[!ht]
    \centering
    \caption{Number of nodes, number of edges and number of elements in each aspect of the model, the latter denoted by $|\cdot|$, according to the study case and whether the MAG was filtered to keep events strictly related to the main pathway of the study case.}
    \label{tab:mag_info}
    \begin{tabular}{cccccccccc}
    \hline
    Study Case & MAG & Filter & Nodes & Edges & $|I|$ & $|D|$ & $|O|$ & $|U|$ & $|S|$ \\ \hline
    Pregnancy & (I,O,U,S) & No & \numprint{561766} & \numprint{3188617} & \numprint{581} & N/A & \numprint{110} & \numprint{940} & \numprint{1751} \\
    Pregnancy & (I,O,U,S) & Yes & \numprint{41650} & \numprint{1248139} & \numprint{53} & N/A & \numprint{44} & \numprint{607} & \numprint{67} \\
    Diabetes & (I,O,U,S) & No & \numprint{470475} & \numprint{1893312} & \numprint{658} & N/A & \numprint{113} & \numprint{941} & \numprint{1316} \\
    Diabetes & (I,D,O,U,S) & No & \numprint{187152} & \numprint{571731} & \numprint{443} & \numprint{4358} & \numprint{77} & \numprint{426} & \numprint{308} \\ \hline
    \end{tabular}
\end{table}

For the pregnancy models, considering the number of elements in each aspect, the most affected ones were Sequence and Intervention, with a decrease of \numprint[\%]{96.17} and \numprint[\%]{90.88}, respectively. The number of occupations dropped by \numprint[\%]{60.0} and the number of healthcare units had a reduction of \numprint[\%]{35.43}. 
In the diabetes case study, the difference in the number of elements in each aspect from the 4-aspect MAG to the 5-aspect one was smaller than those observed in the pregnancy case (\numprint[\%]{32.67}, \numprint[\%]{31.86} and \numprint[\%]{76.60} for aspects $I,O,S$, respectively), except for the aspect Healthcare Unit, which registered a difference of \numprint[\%]{54.73}. One hypothesis that could explain this marked reduction in the number of units is that, because of different organisational cultures, some units might be more likely to register valid diagnosis for each patient encounter than others.

\subsection{Clustering patient pathways}

With the built MAG models, we proceed to the clustering step to group similar pathways before mining the pathway models.
We extracted a sample of \numprint{2000} patient pathways with length less than or equal to the third quartile of pathway length from each of the four MAGs to calculate the dissimilarity between them and cluster the pathways. The virtual events ``start'' and ``end'' do not take part in the pathway comparison.
We took a sample and established a maximum length of patient pathways due to the cost of computing the dissimilarities and clustering the pathways.
We normalised the dissimilarity between two patient pathways dividing it by the sum of the lengths of both pathways.
We adopted the $Dist_A$ defined in Equation~\ref{eq:distA_1} with $\delta=0.5$ for the three 4-aspect MAGs
to forbid alignments between activity tuples whose interventions are different, independent of their occupation. Mind that any $\delta$ value between \numprint{0.3} and \numprint{0.7} would yield the same result. For the diabetes 5-aspect MAG, we used $Dist_A$ as defined in Equation~\ref{eq:distA_2} with $\delta=0.6$ to forbid alignments between activity tuples whose diagnoses do not have the first three digits in common, independent of their intervention. Mind that any $\delta$ value between \numprint{0.51} and \numprint{0.7} would yield the same result.
For all MAGs, we used $\varepsilon = 20$~days.

The  cophenetic correlation coefficient~(CCC) measures how well the dendrogram resulting from an agglomerative hierarchical clustering algorithm can represent the structure of the data. The closer its value is to $1$, the best the clusters reflect the original distances.
We calculate the CCC value using single, complete and average linkage for each dissimilarity matrix to assess if this traditional hard clustering method would fit our patient pathway analysis.
The average linkage returned the best CCC values for every matrix, so we present its results in Table~\ref{tab:cophenetic}. Besides the CCC of the whole matrix ($2000\times2000$), we randomly extracted $25$ submatrices of size $1500\times1500$ from it and calculated the CCC for each one. The mean and standard deviation values are also available in Table~\ref{tab:cophenetic}.

\begin{table}[!ht]
\caption{Cophenetic correlation coefficient (CCC) for hierarchical clustering with average linkage according to the study case and whether the MAG was filtered to keep events strictly related to the main pathway of the study case. For each of the settings, the \textbf{CCC} column refers to the task of clustering a sample of \numprint{2000} pathways and the \textbf{Sampled CCC} column refers to the mean and standard-deviation of the CCC calculated on 25 samples of $1500$ pathways.}
\label{tab:cophenetic}
\nprounddigits{3}
\centering
\begin{threeparttable}[b]
    \begin{tabular}{cccccc}
    \hline
    Study Case & MAG & Filter & CCC & Sampled CCC & Maximum Pathway Length \\ \hline
    Pregnancy  & (I,O,U,S)   & Yes    & \numprint{0.768680853774646} & \numprint{0.782187878628871}$\pm$\numprint{0.00939250643605899} & 11\\
    Pregnancy  & (I,O,U,S)   & No     & \numprint{0.738834633} & \numprint{0.766732681417593}$\pm$\numprint{0.00982911679669869} & 30\\
    Diabetes   & (I,D,O,U,S) & No     & \numprint{0.664633790606181} & \numprint{0.697289557593851}$\pm$\numprint{0.0114422355201428} & 9\\
    Diabetes   & (I,O,U,S)   & No     & \numprint{0.615044587415316}  & \numprint{0.645862527368978}$\pm$\numprint{0.0111860428908273} & 33\\
    \hline
    \end{tabular}
    % \begin{tablenotes}
    % \raggedleft
    %   \small
    %   \item CCC: cophenetic correlation coefficient.
    % \end{tablenotes}
\end{threeparttable}
\end{table}
\npnoround

The higher CCC corresponds to the matrix of the filtered pregnancy pathways, which is the scenario that minimises the variability coming from possibly overlapping pathways. The unfiltered pregnancy case has, in general, a slightly smaller CCC, despite having significantly larger pathways. However, when we move on to the diabetes cases, there is a further decrease in the CCC value, especially in the 4-aspect MAG.
%Although the first $3$ correlations are not extremely close to $1$, as they are greater than \numprint{0.7}, we could admit to cluster them using the hierarchical clustering.
These results reinforce the idea that the higher the variability of patient pathways, the harder it is to group patients using a hard clustering approach. 
Thus, we used the OSLOM algorithm to cluster the pathways.
%Although in the first $3$ cases the correlation in higher than \numprint{0.7}, as they were not extremely close to $1$, we to use the soft clustering approach with them as well.

We set the number of runs of the algorithm to $25$ and repeated the clustering with $5$ different seeds. We chose the result from the seed which led to the lowest mean probability of finding alike clusters in a random network. Table~\ref{tab:resultados_oslom} presents the number of clusters and singletons (pathways that did not fit any cluster) for each case.

\begin{table}[!ht]
    \centering
    \caption{Number of clusters and singletons for each case study.}
    \begin{tabular}{ccccc}
    \hline
        Study Case & MAG & Filter & Number of Clusters & Number of Singletons \\ \hline
        Pregnancy & (I,O,U,S) & Yes & 2 & 614 \\
        Pregnancy & (I,O,U,S) & No & 5 & 313 \\
        Diabetes & (I,O,U,S) & No & 4 & 331 \\
        Diabetes & (I,D,O,U,S) & No & 4 & 4 \\ \hline
    \end{tabular}
    \label{tab:resultados_oslom}
\end{table}

Although the criteria for calculating the dissimilarities among the pathways take into account the sequence of events and the elapsed time.we provide the frequency of each medical procedure and occupation (or medical procedure and diagnosis, for the 5-aspect diabetes MAG) to provide a first evaluation of the clusters. 

The two clusters of the filtered pregnancy case study share the same set of the seven most frequent pairs of intervention and occupation, with similar relative frequencies (see Table~\ref{tab:clusters_gestacao_filtro}). Their main combinations are the antenatal care visit with an obstetrician, followed by antenatal care visits with nurses and doctors from the Family Strategy Program. 
Cluster $1$ is the largest one, encompassing more than half of the patients. 
Despite the similar frequencies, the mean pathway length is greater in Cluster $1$ than in Cluster~$0$ (\numprint{8.7}$\pm$\numprint{1.7} versus \numprint{5.2}$\pm$\numprint{1.8}).

\begin{table}[!ht]
    \centering
    \caption{
    Relative frequency of pairs of interventions and occupations in the clusters of the pregnancy case study (filtered for events directly related to pregnancy.) The seven most frequent combinations of each cluster are displayed. $N_{p}$ expresses the number of patients in the cluster and $\mu$ is the average pathway length.
    }
    \resizebox{0.9\textwidth}{!}{%
\begin{tabular}{@{}
>{\columncolor[HTML]{FFFFFF}}l 
>{\columncolor[HTML]{FFFFFF}}l rrrr@{}}
\toprule
\multicolumn{1}{c}{} & \multicolumn{1}{c}{} & \textbf{Cluster 0} & \textbf{Cluster 1} \\
\multicolumn{1}{c}{} & \multicolumn{1}{c}{} & $N_P = 584$ & $N_P = 1076$ \\
\multicolumn{1}{c}{\multirow{-3}{*}{\textbf{Intervention}}} & \multicolumn{1}{c}{\multirow{-3}{*}{\textbf{Occupation}}} & $\mu=5.2\pm 1.8$ & $\mu=8.7\pm 1.7$ \\ \hline
Antenatal Care Visit & Doctor (Family Health Strategy) & \cellcolor[HTML]{DDECD4}16.52\% & \cellcolor[HTML]{DDECD5}16.21\% \\
Obstetric Ultrassound Scan & Doctor in radiology & \cellcolor[HTML]{F3F7F2}5.89\% & \cellcolor[HTML]{EEF5EC}7.93\% \\
Antenatal Care Visit & Nurse (Family Health Strategy) & \cellcolor[HTML]{DAEAD1}17.90\% & \cellcolor[HTML]{DEECD6}16.00\% \\
Antenatal Care Visit & Doctor in Obstetrics and Gynaecology & \cellcolor[HTML]{A9D08E}41.69\% & \cellcolor[HTML]{ADD293}40.13\% \\
Antenatal Care Visit & Nurse & \cellcolor[HTML]{EAF3E7}10.00\% & \cellcolor[HTML]{EAF3E7}9.98\% \\
Pregnancy test & Nursing Assistant & \cellcolor[HTML]{FCFCFF}1.02\% & \cellcolor[HTML]{FBFCFE}1.62\% \\
Postnatal Care Visit & Doctor in Obstetrics and Gynaecology & \cellcolor[HTML]{FAFBFD}2.04\% & \cellcolor[HTML]{FBFBFD}1.95\% \\ \bottomrule
\end{tabular}
    }  
\label{tab:clusters_gestacao_filtro}
\end{table}

In the unfiltered pregnancy case study, there are $5$ clusters---three more than in the filtered case. There is also a larger variability of the most frequent combinations of intervention and occupation among the clusters (see Table~\ref{tab:clusters_gestacao_completo} for the $5$ most frequent pairs in each group). 
However, the largest cluster (Cluster $4$ with \numprint{1051} patients) resembles the largest cluster of the filtered case. Its leading combination is the antenatal care visit with an obstetrician, but antenatal care visits with nurses and doctors from the Family Health Strategy program and unscheduled primary care visits are also frequent. 
Although Clusters $4$ and $2$ share similar frequencies, the mean pathway length is quite different---the pathways in Cluster $4$ are almost three times longer than those in Cluster $2$ on average.
Cluster $3$ is notably marked by unscheduled primary care visits with a general practitioner, and Cluster $0$ has a high participation of domiciliary visits by community health workers. These two combinations refer both to interventions and occupations that are not specific to pregnancy. The filter used to obtain the pregnancy-specific events did not consider any of them as relevant, so they do not take part in the filtered MAG (unless they had a valid diagnosis associated with them and it was a pregnancy one). 
The three most frequent pairs of Cluster $1$ are primary care visits with doctors and nurses from the Family Health Strategy Program and unscheduled primary care visits with general practitioners; once again, these combinations could not be easily distinguished as pregnancy-related unless the diagnosis information was available. 

\begin{table}[!ht]
    \centering
    \caption{Relative frequency of pairs of interventions and occupations in the clusters of the pregnancy case study (without filter). The four most frequent combinations of each cluster are displayed. $N_{p}$ expresses the number of patients in the cluster and $\mu$ is the average pathway length.}
    \resizebox{\textwidth}{!}{%
\begin{tabular}{@{}
>{\columncolor[HTML]{FFFFFF}}l 
>{\columncolor[HTML]{FFFFFF}}l rrrrr@{}}
\toprule
\multicolumn{1}{c}{} & \multicolumn{1}{c}{} & \textbf{Cluster 0} & \textbf{Cluster 1} & \textbf{Cluster 2} & \textbf{Cluster 3} & \textbf{Cluster 4} \\
\multicolumn{1}{c}{} & \multicolumn{1}{c}{} & $N_P=97$ & $N_P=267$ & $N_P=299$ & $N_P=166$ & $N=1051$ \\
\multicolumn{1}{c}{\multirow{-3}{*}{\textbf{Intervention}}} & \multicolumn{1}{c}{\multirow{-3}{*}{\textbf{Occupation}}} & \multicolumn{1}{l}{$\mu=23.3\pm   5.6$} & \multicolumn{1}{l}{$\mu=12.6\pm 6.5$} & \multicolumn{1}{l}{$\mu=5.2\pm 2.1$} & \multicolumn{1}{l}{$\mu=14.5\pm 6.3$} & \multicolumn{1}{l}{$\mu=17.7\pm 6.2$} \\ \hline
Unscheduled Primary Care Visit & General practitioner & \cellcolor[HTML]{ECF4E9}9.03\% & \cellcolor[HTML]{E2EFDC}14.49\% & \cellcolor[HTML]{DFEDD8}15.98\% & \cellcolor[HTML]{B3D59B}40.17\% & \cellcolor[HTML]{EDF4EB}8.47\% \\
Antenatal Care Visit & Nurse (Family Health Strategy) & \cellcolor[HTML]{F1F6EF}6.60\% & \cellcolor[HTML]{F7FAF8}3.01\% & \cellcolor[HTML]{F3F8F3}5.20\% & \cellcolor[HTML]{F8FAFA}2.40\% & \cellcolor[HTML]{EEF5EB}8.25\% \\
Domiciliary visit & Community health worker & \cellcolor[HTML]{A9D08E}45.29\% & \cellcolor[HTML]{FCFCFE}0.66\% & \cellcolor[HTML]{FCFCFF}0.13\% & \cellcolor[HTML]{FCFCFE}0.58\% & \cellcolor[HTML]{FBFCFE}0.72\% \\
Antenatal Care Visit & Nurse & \cellcolor[HTML]{FCFCFF}0.35\% & \cellcolor[HTML]{FBFCFE}0.86\% & \cellcolor[HTML]{F4F8F4}4.75\% & \cellcolor[HTML]{FCFCFF}0.41\% & \cellcolor[HTML]{F5F8F5}4.43\% \\
Primary Care Visit & Doctor (Family Health Strategy) & \cellcolor[HTML]{F3F7F3}5.31\% & \cellcolor[HTML]{DBEBD2}18.49\% & \cellcolor[HTML]{F6F9F7}3.40\% & \cellcolor[HTML]{EAF2E6}10.45\% & \cellcolor[HTML]{F3F7F2}5.46\% \\
Primary Care Visit (except by physician) & Nurse (Family Health Strategy) & \cellcolor[HTML]{F3F7F2}5.49\% & \cellcolor[HTML]{E9F2E5}10.67\% & \cellcolor[HTML]{F7F9F8}3.27\% & \cellcolor[HTML]{F1F7F0}6.18\% & \cellcolor[HTML]{F2F7F2}5.60\% \\
Antenatal Care Visit & Doctor (Family Health Strategy) & \cellcolor[HTML]{F0F6EE}7.04\% & \cellcolor[HTML]{F6F9F6}3.91\% & \cellcolor[HTML]{F6F9F7}3.53\% & \cellcolor[HTML]{F6F9F6}3.90\% & \cellcolor[HTML]{EDF4EA}8.64\% \\
Antenatal Care Visit & Doctor in Obstetrics and Gynaecology & \cellcolor[HTML]{F6F9F7}3.72\% & \cellcolor[HTML]{F7F9F8}3.22\% & \cellcolor[HTML]{C4DFB3}30.68\% & \cellcolor[HTML]{F5F8F5}4.48\% & \cellcolor[HTML]{D0E5C2}24.58\% \\
Primary Care Visit & General practitioner & \cellcolor[HTML]{FCFCFF}0.13\% & \cellcolor[HTML]{F3F7F2}5.52\% & \cellcolor[HTML]{FAFBFC}1.67\% & \cellcolor[HTML]{FAFBFC}1.45\% & \cellcolor[HTML]{FCFCFE}0.62\% \\
Primary Care Visit & Doctor in Obstetrics and Gynaecology & \cellcolor[HTML]{FBFCFD}1.02\% & \cellcolor[HTML]{EAF3E6}10.23\% & \cellcolor[HTML]{EDF4EB}8.41\% & \cellcolor[HTML]{F7F9F8}3.23\% & \cellcolor[HTML]{F3F8F3}5.19\% \\  \bottomrule
\end{tabular}
    }
\label{tab:clusters_gestacao_completo}
\end{table}

The comparison of the clustering results with and without filter for pregnancy-related events reveals that the largest clusters in each case share similar characteristics, such as the antenatal care visits performed by three different occupations and the performance of obstetric ultrasound scans. Another similar point is that both the filtered and unfiltered cases have a cluster with shorter pathways (Cluster $0$ in the filtered case and Cluster $2$ in the unfiltered one).
The comparison also stresses the usefulness of having the diagnosis field properly filled. Several medical procedures not strictly related to any health condition stood out in the unfiltered case, and we cannot be sure if these are visits related to pregnancy or the result of overlapping pathways. 

The $5$ most frequent combinations of intervention and occupation for each of the five clusters of the diabetes case study (4-aspect MAG) are displayed in Table~\ref{tab:clusters_diabetes_4aspectos}.
Cluster $3$ is the largest one, with \numprint[\%]{46.85} of the patients belonging to it, but it is also the one with the shortest pathways. It is mainly characterised by primary care visits, specially by doctors (general practitioners and doctors of the Family Health Strategy program). Cluster $2$ stands out as the one with the highest participation of secondary care visits with endocrinologists and blood pressure measurements, either by nursing assistants or technicians. The relative frequency of these interventions are at least $9$ times greater in Cluster $2$ than in the remaining ones.
Cluster $0$ has a distinctive share (\numprint[\%]{41.43}) of unscheduled primary care visits with general practitioners, followed by other primary care visits.
Lastly, the top combination of intervention and occupation for Cluster $1$ is the primary care visit with doctors from the Family Health Strategy Program, but its most evident difference from the other clusters is the high frequency of domiciliary visits. 

\begin{table}[!ht]
\centering
\caption{Relative frequency of pairs of interventions and occupations in the clusters of the diabetes case study (4-aspect MAG). The five most frequent combinations of each cluster are displayed. $N_{p}$ expresses the number of patients in the cluster and $\mu$ is the average pathway length.}
\resizebox{\textwidth}{!}{%
\begin{tabular}{@{}
>{\columncolor[HTML]{FFFFFF}}l 
>{\columncolor[HTML]{FFFFFF}}l rrrr@{}}
\toprule
\multicolumn{1}{c}{} & \multicolumn{1}{c}{} & \textbf{Cluster 0} & \textbf{Cluster 1} & \textbf{Cluster 2} & \textbf{Cluster 3} \\
\multicolumn{1}{c}{} & \multicolumn{1}{c}{} & $N_P=214$ & $N_P=213$ & $N_P=402$ & $N_P=937$ \\
\multicolumn{1}{c}{\multirow{-3}{*}{\textbf{Intervention}}} & \multicolumn{1}{c}{\multirow{-3}{*}{\textbf{Occupation}}} & $\mu=13.1\pm 7.4$ & $\mu=20.8\pm 6.2$ & $\mu=15.8\pm 7.9$ & $\mu=8.7\pm 4.7$ \\ \hline
Primary Care Visit & General practitioner & \cellcolor[HTML]{E4F0DF}12.00\% & \cellcolor[HTML]{FCFCFF}0.32\% & \cellcolor[HTML]{DFEDD7}14.82\% & \cellcolor[HTML]{C2DDB0}29.19\% \\
Unscheduled Primary Care Visit & General practitioner & \cellcolor[HTML]{A9D08E}41.43\% & \cellcolor[HTML]{F0F6EF}6.17\% & \cellcolor[HTML]{F3F7F3}4.71\% & \cellcolor[HTML]{F3F7F2}4.91\% \\
Primary Care Visit & Doctor (Family Health Strategy) & \cellcolor[HTML]{ECF4E9}8.29\% & \cellcolor[HTML]{C3DEB1}28.72\% & \cellcolor[HTML]{F8FAFA}2.14\% & \cellcolor[HTML]{CCE3BE}24.14\% \\
Primary Care Visit (except by physician) & Nurse (Family Health Strategy) & \cellcolor[HTML]{F7FAF8}2.64\% & \cellcolor[HTML]{EEF5EC}7.28\% & \cellcolor[HTML]{FAFBFC}1.18\% & \cellcolor[HTML]{F4F8F3}4.41\% \\
Domiciliary assistance & Nursing Assistant (Family Health Strategy) & \cellcolor[HTML]{FCFCFF}0.04\% & \cellcolor[HTML]{F6F9F7}3.12\% & \cellcolor[HTML]{FCFCFF}0.05\% & \cellcolor[HTML]{F9FBFB}1.65\% \\
Blood Pressure Measurement & Nursing Assistant & \cellcolor[HTML]{FCFCFE}0.43\% & \cellcolor[HTML]{FCFCFF}0.14\% & \cellcolor[HTML]{ECF4EA}8.06\% & \cellcolor[HTML]{FBFCFD}0.79\% \\
Secondary Care Visit & Doctor in endocrinology and metabolism & \cellcolor[HTML]{FAFBFC}1.43\% & \cellcolor[HTML]{FBFCFD}0.88\% & \cellcolor[HTML]{E2EEDB}13.23\% & \cellcolor[HTML]{FBFCFE}0.68\% \\
Domiciliary visit & Community health worker & \cellcolor[HTML]{FAFBFC}1.21\% & \cellcolor[HTML]{C8E0B8}26.39\% & \cellcolor[HTML]{FCFCFF}0.08\% & \cellcolor[HTML]{F7FAF8}2.78\% \\
Blood Pressure Measurement & Nursing technician & \cellcolor[HTML]{FCFCFF}0.18\% & \cellcolor[HTML]{FCFCFF}0.00\% & \cellcolor[HTML]{F1F7F0}5.51\% & \cellcolor[HTML]{FCFCFF}0.15\% \\
Primary Care Visit (except by physician) & Social worker & \cellcolor[HTML]{F8FAF9}2.32\% & \cellcolor[HTML]{FCFCFF}0.14\% & \cellcolor[HTML]{FAFBFC}1.10\% & \cellcolor[HTML]{FCFCFE}0.43\% \\ \bottomrule
\end{tabular}
}
\label{tab:clusters_diabetes_4aspectos}
\end{table}

Table~\ref{tab:clusters_diabetes_5_aspectos} presents the three most frequent combinations of intervention and diagnosis for each of the six clusters of the diabetes case study (5-aspect MAG).
Cluster~$3$ is the largest one with \numprint[\%]{75.4} of the patients, and is mainly characterised by secondary care visits due to Type 2 diabetes and hypothyroidism. 
Primary and secondary care visits due to Type 1 diabetes are the main characteristic of Cluster~$0$, whereas in Cluster~$2$, the diabetes type is frequently not recorded.
Primary care visits associated with essential hypertension constitute the main combination of intervention and diagnosis of Cluster~$1$. 
%Most patients (\numprint{1878}) take part in only one cluster and \numprint{122} take part in two clusters.

\begin{table}[!ht]
    \centering
    \caption{Relative frequency of pairs of interventions and diagnosis in the clusters of the diabetes case study (5-aspect MAG). The two most frequent combinations of each cluster are displayed. $N_{p}$ expresses the number of patients in the cluster and $\mu$ is the average pathway length.}
\resizebox{\textwidth}{!}{%
\begin{tabular}{llrrrrrr}
\hline
\rowcolor[HTML]{FFFFFF} 
\multicolumn{1}{c}{} & \multicolumn{1}{c}{} & \textbf{Cluster 0} & \textbf{Cluster 1} & \textbf{Cluster 2} & \textbf{Cluster 3} \\
\multicolumn{1}{c}{} & \multicolumn{1}{c}{} & $N_P=87$ & $N_P=242$ & $N_P=281$ & $N_P=1508$ \\
\multicolumn{1}{c}{\multirow{-3}{*}{\textbf{Intervention}}} & \multicolumn{1}{c}{\multirow{-3}{*}{\textbf{Diagnosis}}} & $\mu=3.3\pm 1.7$ & $\mu=4.8\pm 2.1$ & $\mu=4.5\pm 2.2$ & $\mu=5.0\pm 2.1$ \\ \hline
Primary Care Visit & Unspecified diabetes mellitus & \cellcolor[HTML]{F7FAF8}2.07\% & \cellcolor[HTML]{DDECD5}12.10\% & \cellcolor[HTML]{BDDBA9}24.31\% & \cellcolor[HTML]{FBFCFD}0.59\% \\
Secondary Care Visit & Unspecified diabetes mellitus & \cellcolor[HTML]{F8FAF9}1.72\% & \cellcolor[HTML]{F8FAF9}1.80\% & \cellcolor[HTML]{DEEDD7}11.57\% & \cellcolor[HTML]{FCFCFF}0.15\% \\
Secondary Care Visit & Type 1 diabetes mellitus & \cellcolor[HTML]{BDDBA9}24.48\% & \cellcolor[HTML]{F9FBFB}1.29\% & \cellcolor[HTML]{FBFCFD}0.63\% & \cellcolor[HTML]{F7F9F7}2.29\% \\
Secondary Care Visit & Hypothyroidism, unspecified & \cellcolor[HTML]{FCFCFF}0.00\% & \cellcolor[HTML]{FCFCFF}0.09\% & \cellcolor[HTML]{FAFBFC}0.87\% & \cellcolor[HTML]{E0EDD9}10.98\% \\
Primary Care Visit & Menopausal and female climacteric states & \cellcolor[HTML]{EBF3E7}6.90\% & \cellcolor[HTML]{F6F9F6}2.58\% & \cellcolor[HTML]{F7F9F7}2.28\% & \cellcolor[HTML]{FCFCFF}0.16\% \\
Primary Care Visit & Type 1 diabetes mellitus & \cellcolor[HTML]{C9E1BA}19.66\% & \cellcolor[HTML]{F8FAF9}1.89\% & \cellcolor[HTML]{FAFBFC}0.94\% & \cellcolor[HTML]{FBFCFE}0.55\% \\
Secondary Care Visit & Type 2 diabetes mellitus without complications & \cellcolor[HTML]{FCFCFF}0.00\% & \cellcolor[HTML]{FCFCFF}0.00\% & \cellcolor[HTML]{FBFCFE}0.47\% & \cellcolor[HTML]{D6E8CB}14.83\% \\
Secondary Care Visit & Type 2 diabetes mellitus & \cellcolor[HTML]{FCFCFF}0.00\% & \cellcolor[HTML]{FCFCFF}0.09\% & \cellcolor[HTML]{FCFCFF}0.24\% & \cellcolor[HTML]{DDECD5}12.14\% \\
Primary Care Visit & Essential (primary) hypertension & \cellcolor[HTML]{FCFCFF}0.00\% & \cellcolor[HTML]{A9D08E}31.93\% & \cellcolor[HTML]{E4EFDE}9.52\% & \cellcolor[HTML]{F8FAFA}1.58\% \\
Secondary Care Visit & Essential (primary) hypertension & \cellcolor[HTML]{FBFCFD}0.69\% & \cellcolor[HTML]{F4F8F5}3.09\% & \cellcolor[HTML]{F7FAF8}2.12\% & \cellcolor[HTML]{F3F7F2}3.68\% \\ \hline
\end{tabular}
}  
\label{tab:clusters_diabetes_5_aspectos}
\end{table}

\subsection{Mining patient pathways}

After clustering the pathways, we prepared the mining method to explore their resulting models. 
First, to calculate $R_0$ (Equation~\ref{eq:relevancia_inicial}) for each activity tuple, we used the subdeterminations of the MAG with all patients, i.e. before clustering. We calculated $R_0$ for each of the four MAGs (the two of the pregnancy study case---filtered and unfiltered---and the two of the diabetes case study---with and without the Diagnosis aspect). 
We used the NetworkX implementation of the closeness, betweenness and PageRank algorithms~\cite{networkx}. To identify the most specialised healthcare units, we set $\alpha = 0.85$ in the PageRank algorithm, which is a common choice for the parameter~\cite{surveyPagerank,bookNewman}, and $\beta = \frac{1}{N}$ for each node, where $N$ is the number of nodes in the network.
We normalised the values of the closeness, betweenness and PageRank centralities so that the values of each one of them would range between $0$ and $1$.
After computing the values of $R_0$, we obtain the final relevance of the nodes by applying the PageRank algorithm to each cluster. Once again, we used the NetworkX implementation of the PageRank algorithm \cite{networkx}.
%To obtain the relevance of the nodes for the pathway models of each cluster, we used the NetworkX implementation of the PageRank algorithm \cite{networkx}.%, and experimentally found that a value of $\alpha$ around $0.2-0.3$ led to reasonable results, according to the interpretation of the mined model by a medical specialist.

\begin{figure}[!ht]
    \centering
    \includegraphics[width=\textwidth]{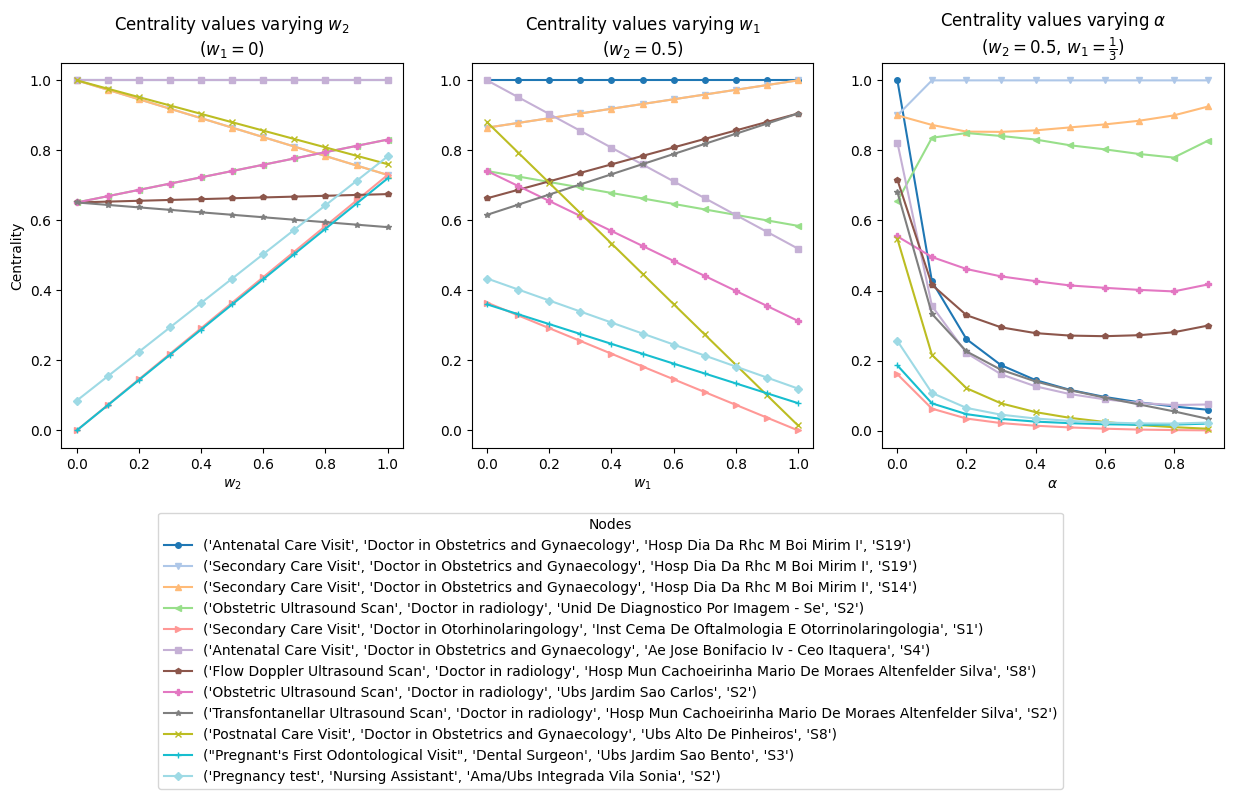}
    \caption{Influence of the parameters $w_1$, $w_2$ and $\alpha$ on the centrality of a sample of nodes from the filtered pregnancy study case.}
    \label{fig:variacao_centralidade}
\end{figure}

The value of $R_0$ depends on the parameters $w_1$ and $w_2$. 
The first weights the relative importance of complications in the pathway, expressed as the PageRank centrality of the healthcare unit in the activity tuple, 
and the second weights the contribution of the closeness value of the intervention and the betweenness value of the occupation to measure the node's relevance as a ``typical event''. 
Besides these parameters, the $\alpha$ parameter of the final PageRank weighs  the influence of the value $R_0$ and the influence of the time and context of the occurrence of each node to its conclusive centrality value. If $\alpha = 0$, only $R_0$ matters to the relevance of the nodes, and as $\alpha$ grows, the influence of $R_0$ diminishes. 
Figure~\ref{fig:variacao_centralidade} illustrates the influence of the three parameters on a set of nodes of the MAG of filtered pregnancy pathways. 
As it would be unfeasible to present all the nodes in the MAG, we selected a sample including nodes whose closeness centrality of intervention was greater, lower and similar than their betweenness centrality of occupation. Similarly, we also included nodes whose ``typical event'' relevance was greater, lower and similar to their ``complication'' relevance. 
In this example, we run the final PageRank in the whole MAG instead of choosing a specific cluster.

The leftmost chart in Figure~\ref{fig:variacao_centralidade} shows how $w_2$ influences the ``typical event'' component of $R_0$. We kept $w_1=0$ and varied $w_2$ from $0$ to $1$. When $w_2 = 0$, only the occupation matters and, when $w_2=1$, only the intervention counts. Thus, the three nodes whose occupation is \textit{Doctor in Obstetrics and Gynaecology} have the same centrality value when $w_2 = 0$, because this is the occupation with the highest betweenness. The node among these three  whose intervention is \textit{Antenatal Care Visit} experiences no change in its centrality because its intervention also has the maximum value of $1$ (highest closeness). On the other hand, the centralities of the other two nodes decrease when $w_2$ grows due to the lower closeness of \textit{Postnatal Care Visit} and \textit{Pregnancy test} when compared to \textit{Antenatal Care Visit}, although both of them are still larger than \numprint{0.5}.  The opposite behaviour occurs for the node whose intervention is \textit{Pregnancy Test} and whose occupation is \textit{Nursing Assistant}; when only the occupation defines the node centrality ($w_2=0$), the node's centrality is very low, close to \numprint{0.1}, but it increases as the intervention participation grows, reaching almost \numprint{0.8}.

The central chart of Figure~\ref{fig:variacao_centralidade} focus on the influence of $w_1$ on the $R_0$ value of the nodes. With $w_2$ fixed as \numprint{0.5}, we varied $w_1$ from $0$ to $1$. When $w_1=0$, only the closeness of the intervention and the betweenness of the occupation affect $R_0$, whereas when $w_1=1$, only the PageRank centrality of the healthcare unit affects $R_0$. Owing to this, the two nodes whose intervention is \textit{Antenatal care visit} and whose occupation is \textit{Doctor in Obstetrics and Gynaecology} have maximum centrality of $1$ when $w_1=0$, but, as $w_1$ increases, only the node that refers to a hospital (\textit{Hosp Dia Da Rhc M Boi Mirim I}) keeps a high centrality value, the other one loses importance. The opposite pattern happens for the nodes whose interventions are the exams \textit{Flow Doppler Ultrasound} and \textit{Transfontanellar Ultrasound scan}---their importance grows as $w_1$ increases.

The chart on the right side of Figure~\ref{fig:variacao_centralidade} reveals the influence of the parameter $\alpha$ of the PageRank algorithm on the final centrality of the nodes. When $\alpha=0$, $R_0$ dominates the result, but as $\alpha$ grows, the influence of the context~(links) of each node increases. An interesting example is the pair of nodes whose intervention is \textit{Secondary Care Visit}; they have the same occupation and healthcare unit, so their curves overlapped completely in the previous two charts. When $alpha=0$ they share the same centrality value once again, but the node of \textit{S19} receives greater importance than the one in  \textit{S14} when $\alpha \neq 0$. As $\alpha$ approaches $1$, several nodes assume very small values while a few nodes have very large values. 

\begin{figure}[!ht]
    \centering
    \frame{\includegraphics[width=0.8\textwidth, trim = 0cm 5cm 0cm 0cm, clip]{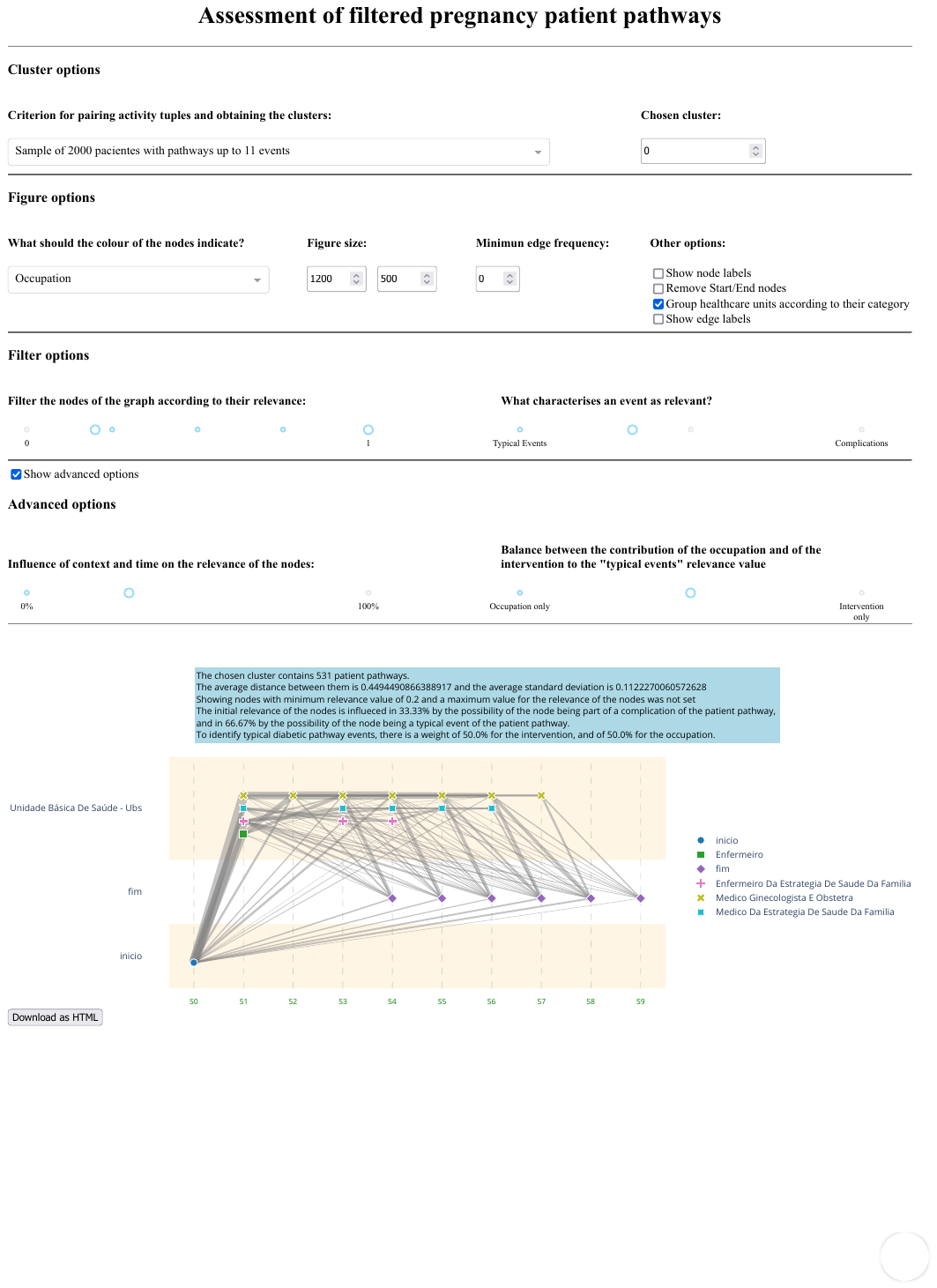}}
    \caption{Tool to aid specialists to select parameters and explore the clusters of pathways.}
    \label{fig:interface_relevance}
\end{figure}

We developed a tool for the specialists to evaluate the best value for the parameters and also to explore the patient pathways of each cluster (Figure~\ref{fig:interface_relevance}). 

After choosing the parameters, each node receives its relevance value as an attribute and it is now possible to remove nodes from the MAG model if their relevance is smaller than a minimum threshold and/or greater than a maximum one. The process of removing a node from the MAG involves linking the nodes that send edges to it to the nodes that receive edges from it. 
Thus, we search for pairs of incoming and outgoing edges with the same patient identifier attribute and create a new edge whose origin is the same as in the incoming edge and whose target is the same as in the outgoing edge. The time interval attribute of the new edge is the sum of the intervals of the two original edges. 
After filtering nodes according to their relevance, some options before generating a visual representation of the model include:
\begin{itemize}
    \item Subdetermine the model to disregard one or more aspects in the visual representation;
    \item Convert the multi-aspect multigraph into a multi-aspect digraph and calculate the frequency of each edge;
    \item Contract nodes according to their attributes, such as the type of healthcare unit;
    \item Hide infrequent edges or start/end nodes;
    \item Colour edges according to the time intervals.
\end{itemize}

\begin{figure}
     \centering
     \begin{subfigure}[b]{\textwidth}
         \centering
         \includegraphics[width=0.75\textwidth]{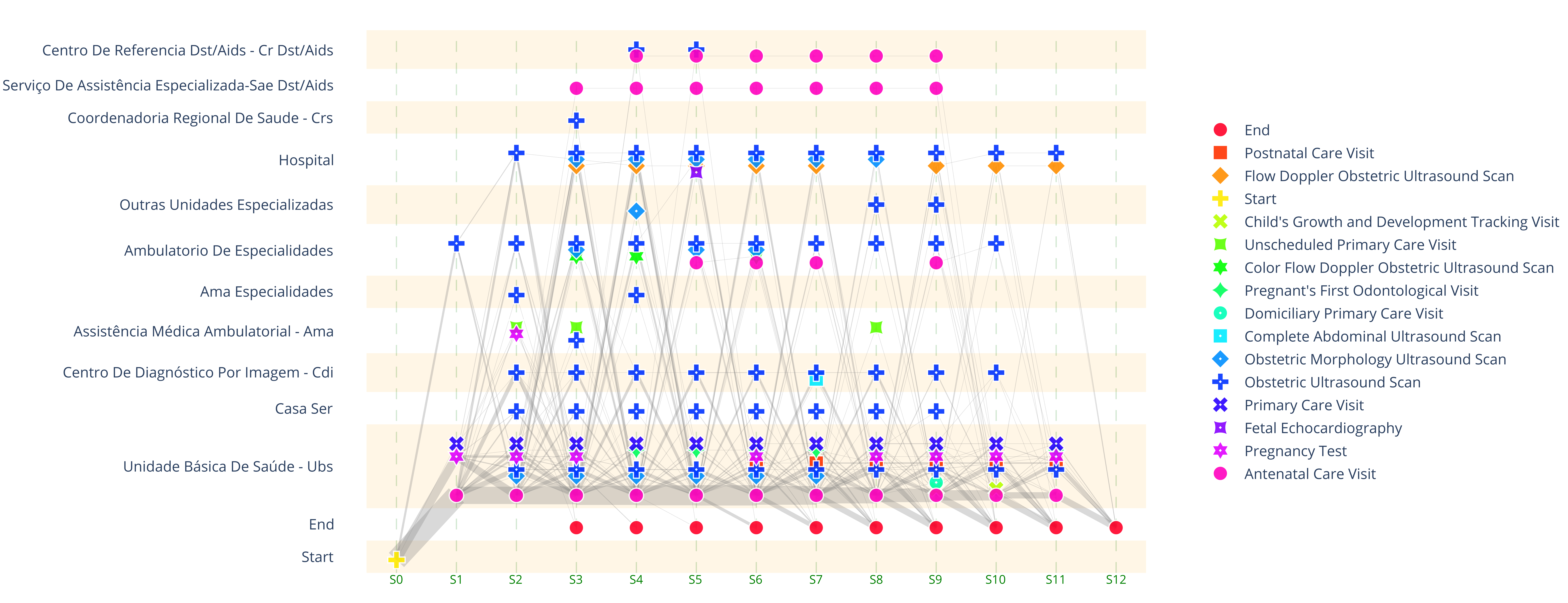}
         \caption{Without filtering nodes according to their relevance.}
         \label{subfig:mag_sem_filtro}
     \end{subfigure}
     \vfill
     \vspace*{5mm}
     \begin{subfigure}[b]{\textwidth}
         \centering
         \includegraphics[width=0.75\textwidth]{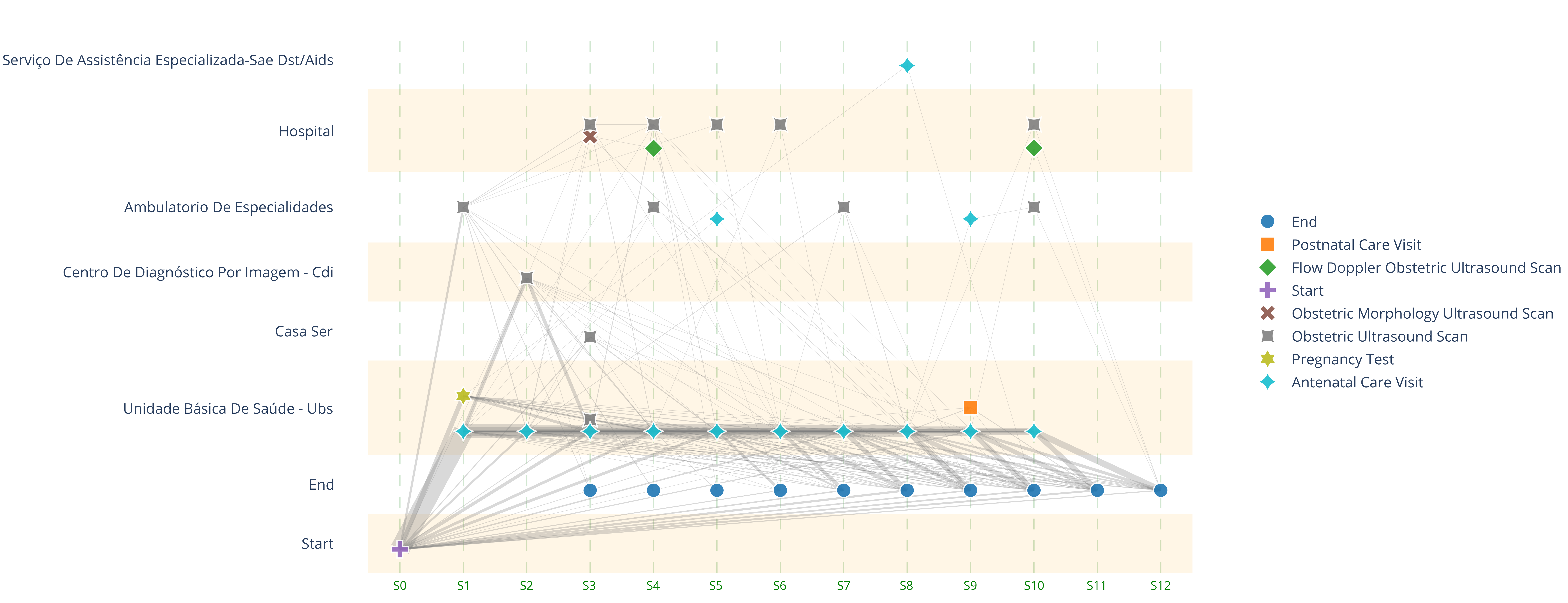}
         \caption{Nodes with relevance greater or equal to $0.4$.}
         \label{subfig:mag_com_filtro}
     \end{subfigure}
     \vfill 
     \vspace*{5mm}
     \begin{subfigure}[b]{\textwidth}
         \centering
         \includegraphics[width=0.7\textwidth]{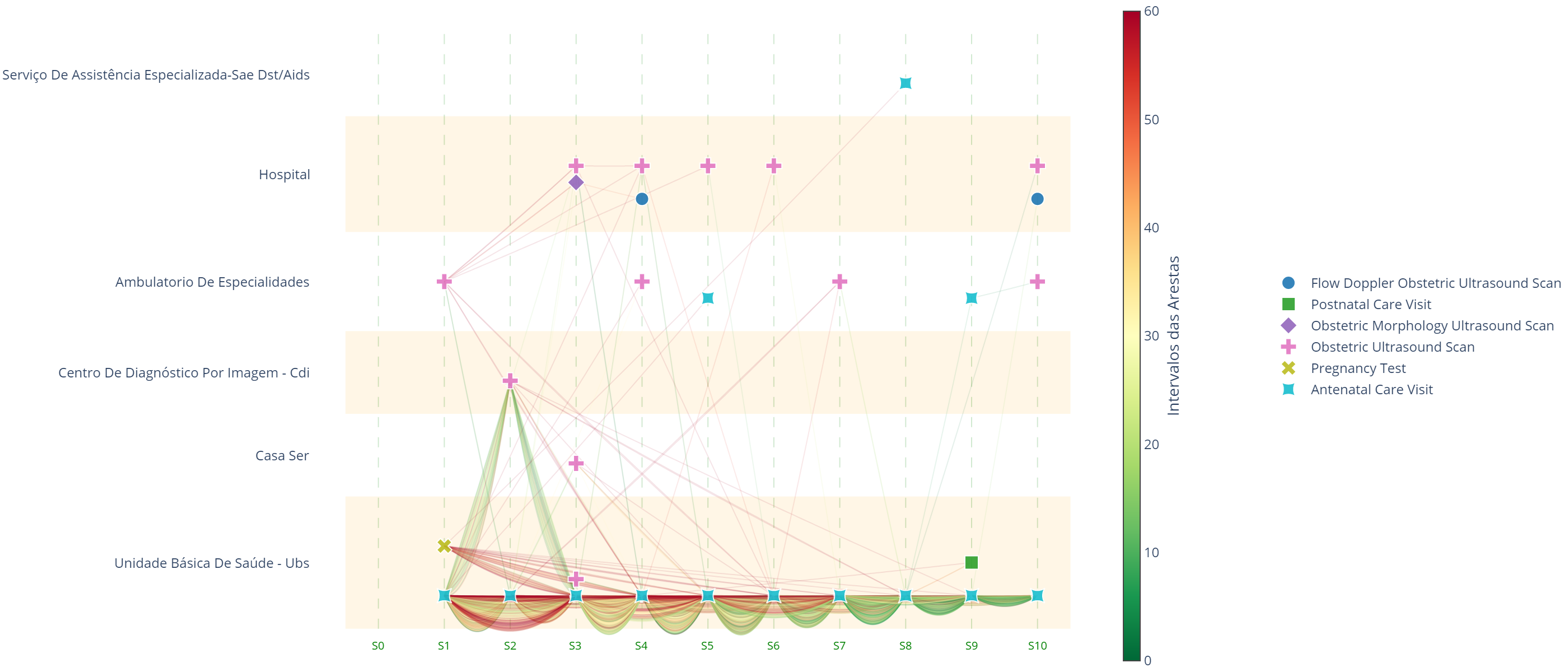}
         \caption{Nodes with relevance greater or equal to $0.4$ and edges coloured according to the time interval. Start and end nodes are not shown.}
         \label{subfig:mag_com_arestas_col}
     \end{subfigure}
     \caption{Visual representation of Cluster $1$ of the filtered pregnancy study case. 
     The stacked horizontal bars represent the categories of healthcare units and the colour/shape of the nodes represent the interventions. The parameters used to calculate the centrality of the nodes were $w_1 = \frac{1}{3}$, $w_2 = 0.5$ and $\alpha = 0.25$.}    
     \label{fig:mags_cluster2}
\end{figure}

%From now, on we adopt $w_1 = \frac{1}{3}$, $w_2 = 0.5$ and $\alpha = 0.25$.
Figure~\ref{fig:mags_cluster2} illustrates Cluster~$2$ of the filtered pregnancy study case. The first chart (Figure~\ref{subfig:mag_sem_filtro}) represents all events of the pathways. The width of the edges is proportional to their frequency. Although it is not a chaotic visual representation, it becomes easier to read it using the filter of nodes. Figure~\ref{subfig:mag_com_filtro} displays the resulting model when only nodes whose centrality is greater or equal to $0.4$ remain in the MAG. 
We see that the main characteristic of this cluster is a repetitive sequence of antenatal care visits in primary care units (\textit{Unidade Básica de Saúde}, in Portuguese), but a pregnancy test as the first event of the pathway and ultrasound scans, especially in $S_1$ and $S_3$ are also important. Some patients in this cluster had antenatal care visits in secondary care units (\textit{Ambulatório de Especialidades}, in Portuguese), which may indicate they had risk factors associated with the pregnancy.
The assessment of the cluster becomes even clearer when we consider the intervals of the edges, as in Figure~\ref{subfig:mag_com_arestas_col}. For instance, the intervals between antenatal care visits seem to become shorter as the pathways approach their end---the first edges range from light green to yellow, while the latter ones have vivid to dark green tones. 

\begin{figure}
     \centering
     \begin{subfigure}[b]{\textwidth}
         \centering
         \includegraphics[width=0.6\textwidth]{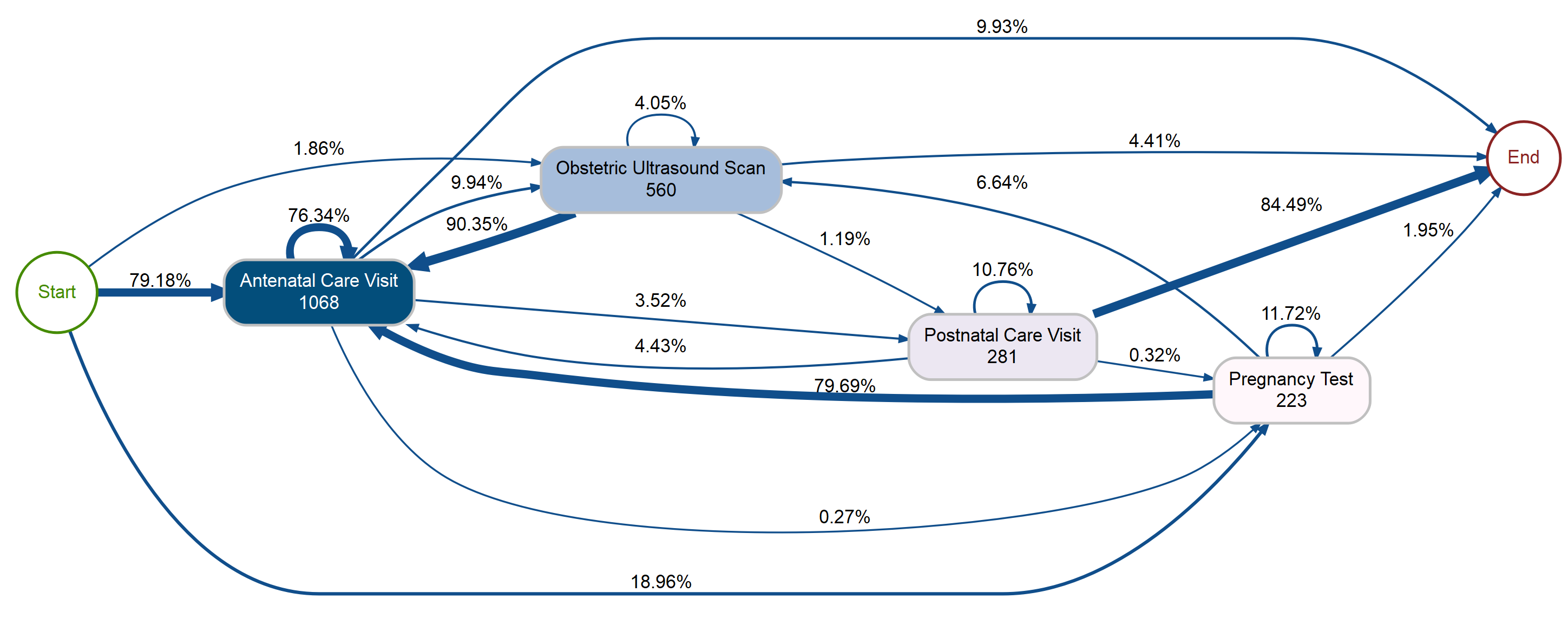}
         \caption{Pregnancy pathway model resulting from the bupaR package. The chart displays only the \numprint[\%]{95} most frequent interventions.}
         \label{subfig:mapa_processo}
     \end{subfigure}
     \vfill
     \begin{subfigure}[b]{\textwidth}
         \centering
         \includegraphics[width=0.8\textwidth]{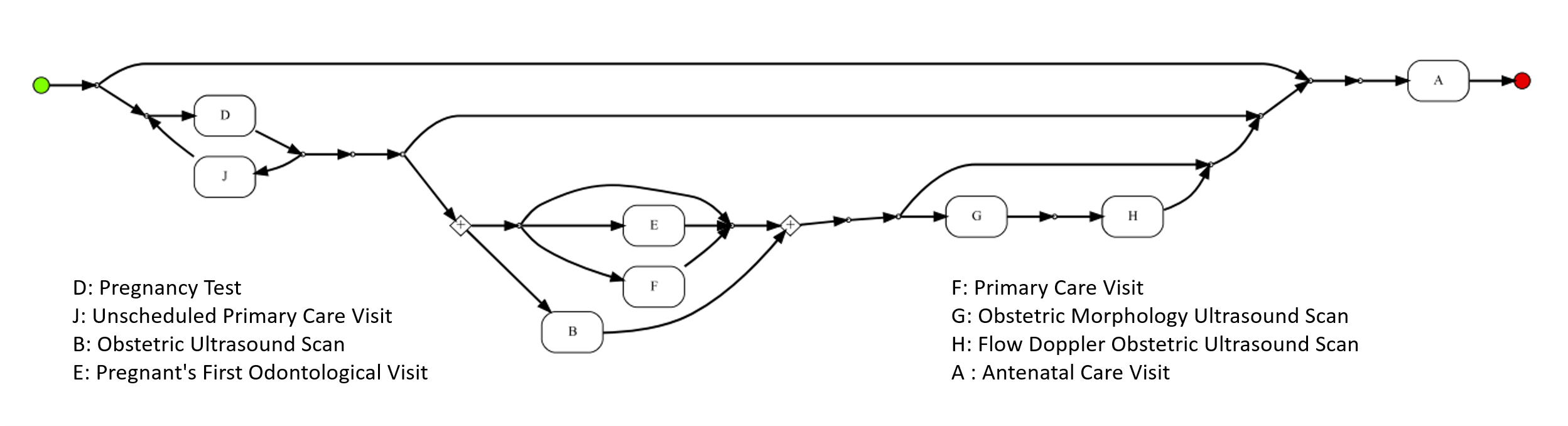}
         \caption{Pregnancy pathway model resulting from the Inductive Miner Infrequent, as implemented in the Inductive Visual Miner plug-in of ProM, with noise threshold set to \numprint{0.4}.}
         \label{subfig:processTree}
     \end{subfigure}
     \caption{Patient pathway models of cluster $1$ of the pregnancy case study using other model/mining methods.}    \label{fig:comparacao_gestantes_cluster2}
\end{figure}

%The MAG graphics allowed us to identify not only the interventions the patients underwent, but also where they happened and in which order. 
As a matter of comparison, we used the same records of the MAG in Figure~\ref{subfig:mag_sem_filtro} to obtain a process map using two well-established and popular Process Mining tools---bupaR~\cite{bupaR} (Figure~\ref{subfig:mapa_processo}) and an Inductive Visual Model with the Inductive Visual Miner plugin~\cite{leemans_inductive} of ProM~\cite{ProM6}~(Figure~\ref{subfig:processTree}). The Inductive Miner guarantees a sound model with perfect fitness, while the process maps obtained with a strategy based on the Fuzzy Miner algorithm are intuitive and flexible to use. However, when it comes to mining multi-perspective patient pathways with significant variability and repetitive actions, their effectiveness is affected. In both graphics of Figure~\ref{fig:comparacao_gestantes_cluster2}, for instance,  it is not possible to know how many times the antenatal care visits actually happen for most patients in the cluster, at which moment of the pathways the patients have an obstetric ultrasound scan, or how long are the intervals between the events through the course of the patient pathways. 
% filtro de frequência

% ProM
% leitura do arquivo csv
% plugin Convert CSV to XES
% plugin Mine Process Tree with Inductive Miner - S.J.J. Leemans

\begin{figure}[!ht]
    \centering
    \includegraphics[width=0.99\textwidth]{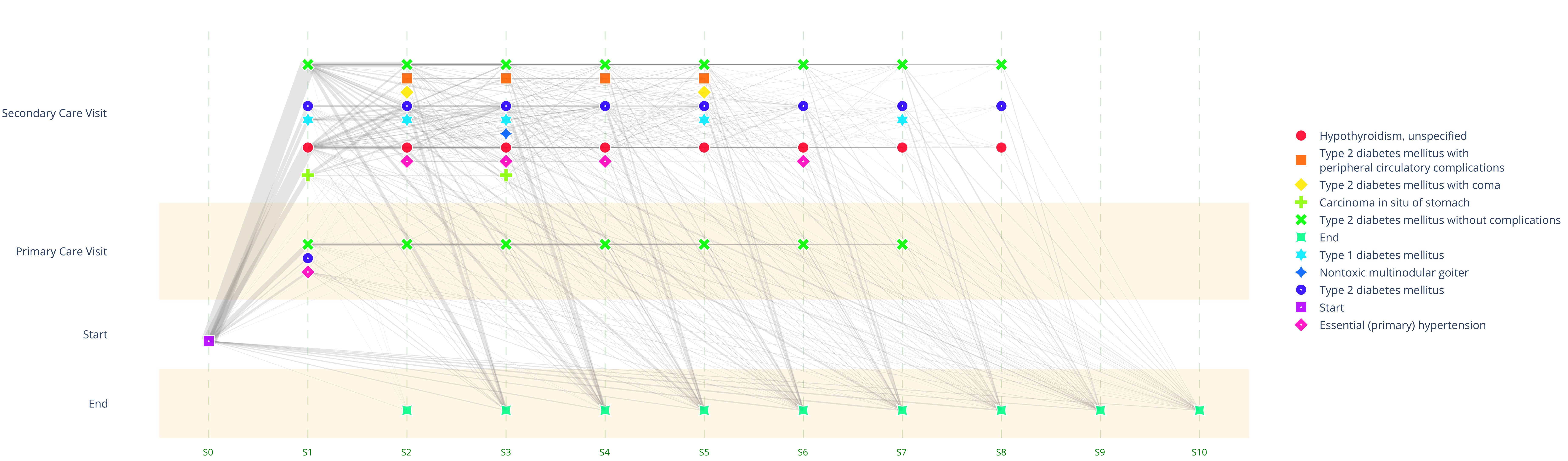}
    \caption{Visual representation of Cluster~$3$ of the diabetic study case (5-aspect MAG). 
    The stacked horizontal bars represent the interventions and the colour/shape of the nodes represent the diagnoses. The parameters used to calculate the centrality of the nodes were $w_1 = 0.5$, $w_2 = 0.5$ and $\alpha = 0.25$. The figure displays only nodes whose centrality is greater than or equal to $0.06$.}
    \label{fig:mag_diabeticos_5asp_cluster4}
\end{figure}

When it comes to the study case of diabetic patients, the mining method allows us to assess the comorbidities patients go through. Figure~\ref{fig:mag_diabeticos_5asp_cluster4} displays the largest cluster~(Cluster~$4$) of the diabetes study case using the 5-aspect MAG. 
Despite the presence of the aspect Diagnosis, we calculated the initial relevance of the nodes~($R_0$) of the 5-aspect MAG using the same strategy used for the 4-aspect MAG, i.e.\ considering only the aspects Intervention, Occupation and Healthcare Unit. 
Secondary care visits are frequent in this group and type~$2$ diabetes is the main diagnosis. Nonetheless, the model also reveals that many patients received the diagnosis of hypothyroidism. There are also records of primary hypertension and diabetes complications such as coma and peripheral circulatory issues.

% ----------------------------------------------
\section{Discussion and Conclusion}
\label{sec:conclusion}

The framework we present in this article successfully helped us to assess multi-perspective patient pathways in two different contexts of high variability and repetitive patterns: pregnancy and diabetes. 
The proposed dissimilarity measurement compares patient pathways taking into account what happened, in which order and with which time intervals, besides allowing the comparison of activities in the pathways in a non-binary way. 
We used this measurement to find clusters of patients with similar behaviour, before discovering the pathway model.
% métrica de similaridade para medir desvio/aderência a um padrão desejável
Although we did not explore it in this work, the dissimilarity measurement could also be a means of evaluating how close/distant a patient's pathway is from an expected or desired standard. Future work could examine this possibility.

%Clustering -- related work
Because of the potential existence of overlapping pathways and of the features of our dissimilarity measurement, we opted to use an overlapping clustering approach to group patients according to their pathways. We used the OSLOM algorithm, which finds overlapping communities in graphs to do so. The use of fuzzy methods to cluster patient pathways is not a novelty. \citeauthor{Prokofyeva2020a} used topic models to find fuzzy clusters of patients with sepsis. Their motivation was to support decision planning based on the identification of different behavioural patterns and the probability of the patients belonging to each one. %not directly the possibility of overlapping pathways, but rather the goal of  
An example of deliberate clustering of patients with multi- and comorbidities is the work of \citeauthor{Najjar2018}. The authors used two clustering steps to study pathways of patients with heart failure disease and found several clusters of patients with comorbidities related to kidney and respiratory problems.

The next step of the framework is mining the patient pathway model. We take advantage of the fact that we can analyse each aspect of the MAG independently of the others to attribute part of the relevance of a node to the relevance of its components to each subdetermination. 
One point worth considering is the choice of $R_0$ when dealing with subsets of patients, such as clusters of pathways. The value of $R_0$ may be the result of the centralities of the whole cohort of patients or only the selected ones. In the former, $R_0$ has the advantage of being more specific to the studied group; however, if this group is too small, there may not be enough data to achieve credible results. For instance, in our study case, we used the PageRank algorithm to find the most specialised healthcare units. If the network of healthcare units were too small, it would be harder to identify such dynamics when compared to taking the whole health system at once.  
We also highlight that the inclusion of the virtual nodes indicating the beginning and the end of the pathways affects the result of the final relevance of the nodes if they are kept for the execution of the Page Rank algorithm. Here, we did not include them in the mining step and only included them in the visual representation of the model.
% escolha de R0 quando há clusterização ou subsets

%\subsection{Framework applicability to other fields of research}
Although we originally designed the framework to model and mine patient pathways, our approach is analogous to the sub-field of Process Mining that deals with discovering unstructured process models~\cite{Gunther2007,Stefanini2020novo}, but focusing on multi-perspective analysis. Unstructured processes are those that are highly dependent on the case being treated and, thus, do not have clearly defined underlying rules to guide their progress---such as the patient pathways we studied in this work. There is no restriction on the applicability of the framework to other fields of research, provided that the centralities of the nodes are adapted to meet the characteristics of the area. 

% 2000
One limitation of this study is that the clustering and mining steps involved a sample of \numprint{2000} pathways instead of the whole cohort. 
We intend to apply the framework to a larger group of patients in future work.
%Another topic left for future research is the investigation of whether it is possible to convert the model resulting from the mining method into a process algebra representation
Another topic left for future research is the investigation of whether the mining method of the framework could work as an auxiliary step to clean the dataset keeping only the most relevant events, before the application of other process discovery algorithms.

% ----------------------------------------------

%\section*{Acknowledgments}

%Bibliography
\bibliography{manuscrito}  

\begin{thebibliography}{54}
\providecommand{\natexlab}[1]{#1}
\providecommand{\url}[1]{\texttt{#1}}
\expandafter\ifx\csname urlstyle\endcsname\relax
  \providecommand{\doi}[1]{doi: #1}\else
  \providecommand{\doi}{doi: \begingroup \urlstyle{rm}\Url}\fi

\bibitem[Andrews et~al.(2020)Andrews, Wynn, Vallmuur, {Ter Hofstede}, and
  Bosley]{Andrews2020}
Robert Andrews, Moe~T. Wynn, Kirsten Vallmuur, Arthur~H.M. {Ter Hofstede}, and
  Emma Bosley.
\newblock {A comparative process mining analysis of road trauma patient
  pathways}.
\newblock \emph{International Journal of Environmental Research and Public
  Health}, 17\penalty0 (10), 5 2020.
\newblock ISSN 16604601.
\newblock \doi{10.3390/ijerph17103426}.

\bibitem[Antonelli et~al.(2012)Antonelli, Baralis, Bruno, Chiusano, Mahoto, and
  Petrigni]{Antonelli2012}
Dario Antonelli, Elena Baralis, Giulia Bruno, Silvia Chiusano, Naeem~A. Mahoto,
  and Caterina Petrigni.
\newblock {Analysis of diagnostic pathways for colon cancer}.
\newblock \emph{Flexible Services and Manufacturing Journal}, 24\penalty0
  (4):\penalty0 379--399, 12 2012.
\newblock ISSN 19366582.
\newblock \doi{10.1007/s10696-011-9095-2}.

\bibitem[Arnolds and Gartner(2018)]{Arnolds2018}
Ines~Verena Arnolds and Daniel Gartner.
\newblock {Improving hospital layout planning through clinical pathway mining}.
\newblock \emph{Annals of Operations Research}, 263\penalty0 (1-2):\penalty0
  453--477, 4 2018.
\newblock ISSN 15729338.
\newblock \doi{10.1007/s10479-017-2485-4}.

\bibitem[Aspland et~al.(2021)Aspland, Harper, Gartner, Webb, and
  Barrett-Lee]{Aspland2021}
Emma Aspland, Paul~R. Harper, Daniel Gartner, Philip Webb, and Peter
  Barrett-Lee.
\newblock {Modified Needleman–Wunsch algorithm for clinical pathway
  clustering}.
\newblock \emph{Journal of Biomedical Informatics}, 115, 3 2021.
\newblock ISSN 15320464.
\newblock \doi{10.1016/j.jbi.2020.103668}.

\bibitem[Basole et~al.(2015)Basole, Braunstein, Kumar, Park, Kahng, Chau,
  Tamersoy, Hirsh, Serban, Bost, Lesnick, Schissel, and Thompson]{Basole2015}
Rahul~C. Basole, Mark~L. Braunstein, Vikas Kumar, Hyunwoo Park, Minsuk Kahng,
  Duen~Horng Chau, Acar Tamersoy, Daniel~A. Hirsh, Nicoleta Serban, James Bost,
  Burton Lesnick, Beth~L. Schissel, and Michael Thompson.
\newblock {Understanding variations in pediatric asthma care processes in the
  emergency department using visual analytics}.
\newblock \emph{Journal of the American Medical Informatics Association},
  22\penalty0 (2):\penalty0 318--323, 3 2015.
\newblock ISSN 1527974X.
\newblock \doi{10.1093/jamia/ocu016}.

\bibitem[Bodenheimer et~al.(2002)Bodenheimer, Lorig, Holman, and
  Grumbach]{PatientSelfManagement}
Thomas Bodenheimer, Kate Lorig, Halsted Holman, and Kevin Grumbach.
\newblock {Patient Self-management of Chronic Disease in Primary Care}.
\newblock \emph{JAMA}, 288\penalty0 (19):\penalty0 2469--2475, 11 2002.
\newblock ISSN 0098-7484.
\newblock \doi{10.1001/jama.288.19.2469}.
\newblock URL \url{https://doi.org/10.1001/jama.288.19.2469}.

\bibitem[Conca et~al.(2018)Conca, Saint-Pierre, Herskovic, Sep{\'{u}}lveda,
  Capurro, Prieto, and Fernandez-Llatas]{Conca2018}
Tania Conca, Cecilia Saint-Pierre, Valeria Herskovic, Marcos Sep{\'{u}}lveda,
  Daniel Capurro, Florencia Prieto, and Carlos Fernandez-Llatas.
\newblock {Multidisciplinary collaboration in the treatment of patients with
  type 2 diabetes in primary care: Analysis using process mining}.
\newblock \emph{Journal of Medical Internet Research}, 20\penalty0 (4), 2018.
\newblock ISSN 14388871.
\newblock \doi{10.2196/jmir.8884}.

\bibitem[Dagliati et~al.(2017)Dagliati, Sacchi, Zambelli, Tibollo, Pavesi,
  Holmes, and Bellazzi]{Dagliati2017}
A.~Dagliati, L.~Sacchi, A.~Zambelli, V.~Tibollo, L.~Pavesi, J.~H. Holmes, and
  R.~Bellazzi.
\newblock {Temporal electronic phenotyping by mining careflows of breast cancer
  patients}.
\newblock \emph{Journal of Biomedical Informatics}, 66:\penalty0 136--147, 2
  2017.
\newblock ISSN 15320464.
\newblock \doi{10.1016/j.jbi.2016.12.012}.

\bibitem[Dagliati et~al.(2018)Dagliati, Tibollo, Cogni, Chiovato, Bellazzi, and
  Sacchi]{Dagliati2018}
Arianna Dagliati, Valentina Tibollo, Giulia Cogni, Luca Chiovato, Riccardo
  Bellazzi, and Lucia Sacchi.
\newblock {Careflow Mining Techniques to Explore Type 2 Diabetes Evolution}.
\newblock \emph{Journal of Diabetes Science and Technology}, 12\penalty0
  (2):\penalty0 251--259, 3 2018.
\newblock ISSN 19322968.
\newblock \doi{10.1177/1932296818761751}.

\bibitem[{De Oliveira} et~al.(2020){De Oliveira}, Augusto, Jouaneton,
  Lamarsalle, Prodel, and Xie]{DeOliveira2020a}
Hugo {De Oliveira}, Vincent Augusto, Baptiste Jouaneton, Ludovic Lamarsalle,
  Martin Prodel, and Xiaolan Xie.
\newblock {Optimal process mining of timed event logs}.
\newblock \emph{Information Sciences}, 528:\penalty0 58--78, 8 2020.
\newblock ISSN 00200255.
\newblock \doi{10.1016/j.ins.2020.04.020}.

\bibitem[de~Oliveira et~al.(2020)de~Oliveira, Prodel, Lamarsalle, Inada-Kim,
  Ajayi, Wilkins, Sekelj, Beecroft, Snow, Slater, and
  Orlowski]{DeOliveira2020b}
Hugo de~Oliveira, Martin Prodel, Ludovic Lamarsalle, Matt Inada-Kim, Kenny
  Ajayi, Julia Wilkins, Sara Sekelj, Sue Beecroft, Sally Snow, Ruth Slater, and
  Andi Orlowski.
\newblock {“Bow-tie” optimal pathway discovery analysis of sepsis hospital
  admissions using the Hospital Episode Statistics database in England}.
\newblock \emph{JAMIA Open}, 3\penalty0 (3):\penalty0 439--448, 2020.
\newblock ISSN 25742531.
\newblock \doi{10.1093/JAMIAOPEN/OOAA039}.

\bibitem[Defossez et~al.(2014)Defossez, Rollet, Dameron, and
  Ingrand]{Defossez2014}
Gautier Defossez, Alexandre Rollet, Olivier Dameron, and Pierre Ingrand.
\newblock {Temporal representation of care trajectories of cancer patients
  using data from a regional information system: An application in breast
  cancer}.
\newblock \emph{BMC Medical Informatics and Decision Making}, 14\penalty0
  (1):\penalty0 24, 12 2014.
\newblock ISSN 14726947.
\newblock \doi{10.1186/1472-6947-14-24}.
\newblock URL
  \url{https://bmcmedinformdecismak.biomedcentral.com/articles/10.1186/1472-6947-14-24}.

\bibitem[Dhaenens et~al.(2018)Dhaenens, Jacques, Vandewalle, Vandromme,
  Chazard, Preda, Amarioarei, Chaiwuttisak, Cozma, Ficheur, Kessaci, Perichon,
  Taillard, Bordet, Lansiaux, Jourdan, Delerue, and Hansske]{Dhaenens2018}
C.~Dhaenens, J.~Jacques, V.~Vandewalle, M.~Vandromme, E.~Chazard, C.~Preda,
  A.~Amarioarei, P.~Chaiwuttisak, C.~Cozma, G.~Ficheur, M.-E. Kessaci,
  R.~Perichon, J.~Taillard, R.~Bordet, A.~Lansiaux, L.~Jourdan, D.~Delerue, and
  A.~Hansske.
\newblock Clinmine: Optimizing the management of patients in hospital.
\newblock \emph{IRBM}, 39\penalty0 (2):\penalty0 83--92, 2018.
\newblock ISSN 1959-0318.
\newblock \doi{https://doi.org/10.1016/j.irbm.2017.12.002}.
\newblock URL
  \url{https://www.sciencedirect.com/science/article/pii/S1959031817301914}.

\bibitem[Duma and Aringhieri(2020)]{Duma2020}
Davide Duma and Roberto Aringhieri.
\newblock {An ad hoc process mining approach to discover patient paths of an
  Emergency Department}.
\newblock \emph{Flexible Services and Manufacturing Journal}, 32\penalty0
  (1):\penalty0 6--34, 3 2020.
\newblock ISSN 19366590.
\newblock \doi{10.1007/s10696-018-9330-1}.

\bibitem[Durojaiye et~al.(2018)Durojaiye, McGeorge, Puett, Stewart, Fackler,
  Hoonakker, Lehmann, and Gurses]{Durojaiye2018}
Ashimiyu~B. Durojaiye, Nicolette~M. McGeorge, Lisa~L. Puett, Dylan Stewart,
  James~C. Fackler, Peter~L.T. Hoonakker, Harold~P. Lehmann, and Ayse~P.
  Gurses.
\newblock {Mapping the Flow of Pediatric Trauma Patients Using Process Mining}.
\newblock \emph{Applied Clinical Informatics}, 9\penalty0 (3):\penalty0
  654--666, 7 2018.
\newblock ISSN 18690327.
\newblock \doi{10.1055/s-0038-1668089}.

\bibitem[Egho et~al.(2014)Egho, Jay, Ra{\"{i}}ssi, Ienco, Poncelet, Teisseire,
  and Napoli]{Egho2014}
Elias Egho, Nicolas Jay, Chedy Ra{\"{i}}ssi, Dino Ienco, Pascal Poncelet,
  Maguelonne Teisseire, and Amedeo Napoli.
\newblock {A contribution to the discovery of multidimensional patterns in
  healthcare trajectories}.
\newblock \emph{Journal of Intelligent Information Systems}, 42\penalty0
  (2):\penalty0 283--305, 4 2014.
\newblock ISSN 15737675.
\newblock \doi{10.1007/s10844-014-0309-4}.

\bibitem[Egho et~al.(2015)Egho, Ra{\"{i}}ssi, Calders, Jay, and
  Napoli]{Egho2015}
Elias Egho, Chedy Ra{\"{i}}ssi, Toon Calders, Nicolas Jay, and Amedeo Napoli.
\newblock {On measuring similarity for sequences of itemsets}.
\newblock \emph{Data Mining and Knowledge Discovery}, 29\penalty0 (3):\penalty0
  732--764, may 2015.
\newblock ISSN 1384-5810.
\newblock \doi{10.1007/s10618-014-0362-1}.
\newblock URL \url{http://link.springer.com/10.1007/s10618-014-0362-1}.

\bibitem[Gonzalez-Garcia et~al.(2020)Gonzalez-Garcia, Telleria-Orriols,
  Estupinan-Romero, and Bernal-Delgado]{Gonzalez-Garcia2020}
Juan Gonzalez-Garcia, Carlos Telleria-Orriols, Francisco Estupinan-Romero, and
  Enrique Bernal-Delgado.
\newblock {Construction of Empirical Care Pathways Process Models from Multiple
  Real-World Datasets}.
\newblock \emph{IEEE Journal of Biomedical and Health Informatics}, 24\penalty0
  (9):\penalty0 2671--2680, 2020.
\newblock ISSN 21682208.
\newblock \doi{10.1109/JBHI.2020.2971146}.

\bibitem[G{\"u}nther and van~der Aalst(2007)]{Gunther2007}
Christian~W. G{\"u}nther and Wil M.~P. van~der Aalst.
\newblock Fuzzy mining -- adaptive process simplification based on
  multi-perspective metrics.
\newblock In Gustavo Alonso, Peter Dadam, and Michael Rosemann, editors,
  \emph{Business Process Management}, pages 328--343, Berlin, Heidelberg, 2007.
  Springer Berlin Heidelberg.
\newblock ISBN 978-3-540-75183-0.

\bibitem[Hagberg et~al.(2008)Hagberg, Schult, and Swart]{networkx}
Aric~A. Hagberg, Daniel~A. Schult, and Pieter~J. Swart.
\newblock Exploring network structure, dynamics, and function using networkx.
\newblock In Ga\"el Varoquaux, Travis Vaught, and Jarrod Millman, editors,
  \emph{Proceedings of the 7th Python in Science Conference}, pages 11 -- 15,
  Pasadena, CA USA, 2008.

\bibitem[Hirschberg(1975)]{hirschberg1975linear}
Daniel~S. Hirschberg.
\newblock A linear space algorithm for computing maximal common subsequences.
\newblock \emph{Communications of the ACM}, 18\penalty0 (6):\penalty0 341--343,
  1975.

\bibitem[Janssenswillen et~al.(2019)Janssenswillen, Depaire, Swennen, Jans, and
  Vanhoof]{bupaR}
Gert Janssenswillen, Benoît Depaire, Marijke Swennen, Mieke Jans, and Koen
  Vanhoof.
\newblock bupar: Enabling reproducible business process analysis.
\newblock \emph{Knowledge-Based Systems}, 163:\penalty0 927--930, 2019.
\newblock ISSN 0950-7051.
\newblock \doi{https://doi.org/10.1016/j.knosys.2018.10.018}.
\newblock URL
  \url{https://www.sciencedirect.com/science/article/pii/S0950705118305045}.

\bibitem[Kempa-Liehr et~al.(2020)Kempa-Liehr, Lin, Britten, Armstrong, Wallace,
  Mordaunt, and O'Sullivan]{Kempa-Liehr2020}
Andreas~W. Kempa-Liehr, Christina Yin~Chieh Lin, Randall Britten, Delwyn
  Armstrong, Jonathan Wallace, Dylan Mordaunt, and Michael O'Sullivan.
\newblock {Healthcare pathway discovery and probabilistic machine learning}.
\newblock \emph{International Journal of Medical Informatics}, 137, 5 2020.
\newblock ISSN 18728243.
\newblock \doi{10.1016/j.ijmedinf.2020.104087}.

\bibitem[Khan et~al.(2018)Khan, Uddin, and Srinivasan]{Khan2018}
Arif Khan, Shahadat Uddin, and Uma Srinivasan.
\newblock {Comorbidity network for chronic disease: A novel approach to
  understand type 2 diabetes progression}.
\newblock \emph{International Journal of Medical Informatics}, 115:\penalty0
  1--9, 7 2018.
\newblock ISSN 18728243.
\newblock \doi{10.1016/j.ijmedinf.2018.04.001}.

\bibitem[Kovalchuk et~al.(2018)Kovalchuk, Funkner, Metsker, and
  Yakovlev]{Kovalchuk2018}
Sergey~V. Kovalchuk, Anastasia~A. Funkner, Oleg~G. Metsker, and Aleksey~N.
  Yakovlev.
\newblock {Simulation of patient flow in multiple healthcare units using
  process and data mining techniques for model identification}.
\newblock \emph{Journal of Biomedical Informatics}, 82:\penalty0 128--142, 6
  2018.
\newblock ISSN 15320464.
\newblock \doi{10.1016/j.jbi.2018.05.004}.

\bibitem[Lancichinetti et~al.(2011)Lancichinetti, Radicchi, Ramasco, and
  Fortunato]{oslom}
Andrea Lancichinetti, Filippo Radicchi, José~J. Ramasco, and Santo Fortunato.
\newblock Finding statistically significant communities in networks.
\newblock \emph{PLOS ONE}, 6\penalty0 (4):\penalty0 1--18, 04 2011.
\newblock \doi{10.1371/journal.pone.0018961}.
\newblock URL \url{https://doi.org/10.1371/journal.pone.0018961}.

\bibitem[Langville and Meyer(2005)]{surveyPagerank}
Amy~N. Langville and Carl~D. Meyer.
\newblock A survey of eigenvector methods for web information retrieval.
\newblock \emph{SIAM Review}, 47\penalty0 (1):\penalty0 135--161, 2005.
\newblock \doi{10.1137/S0036144503424786}.
\newblock URL \url{https://doi.org/10.1137/S0036144503424786}.

\bibitem[Leemans et~al.(2014)Leemans, Fahland, and Van
  Der~Aalst]{leemans_inductive}
Sander~JJ Leemans, Dirk Fahland, and Wil~MP Van Der~Aalst.
\newblock Process and deviation exploration with inductive visual miner.
\newblock In \emph{12th International Conference on Business Process
  Management, BPM 2014}, pages 46--50. CEUR-WS. org, 2014.

\bibitem[Levenshtein et~al.(1966)]{levenshtein1966binary}
Vladimir~I Levenshtein et~al.
\newblock Binary codes capable of correcting deletions, insertions, and
  reversals.
\newblock In \emph{Soviet physics doklady}, volume~10, pages 707--710. Soviet
  Union, 1966.

\bibitem[Manktelow et~al.(2022)Manktelow, Iftikhar, Bucholc, McCann, and
  O'Kane]{Manktelow2022}
Matthew Manktelow, Aleeha Iftikhar, Magda Bucholc, Michael McCann, and Maurice
  O'Kane.
\newblock {Clinical and operational insights from data-driven care pathway
  mapping: a systematic review}.
\newblock \emph{BMC Medical Informatics and Decision Making}, 22\penalty0
  (1):\penalty0 43, dec 2022.
\newblock ISSN 1472-6947.
\newblock \doi{10.1186/s12911-022-01756-2}.
\newblock URL \url{https://doi.org/10.1186/s12911-022-01756-2
  https://bmcmedinformdecismak.biomedcentral.com/articles/10.1186/s12911-022-01756-2}.

\bibitem[Mendes(2018)]{Mendes2018}
Eugênio~Vilaça Mendes.
\newblock The care for chronic conditions in primary health care.
\newblock \emph{Brazilian Journal in Health Promotion}, 31\penalty0 (2), Jun.
  2018.
\newblock \doi{10.5020/18061230.2018.7839}.
\newblock URL \url{https://ojs.unifor.br/RBPS/article/view/7839}.

\bibitem[Najjar et~al.(2018)Najjar, Reinharz, Girouard, and
  Gagn{\'{e}}]{Najjar2018}
Ahmed Najjar, Daniel Reinharz, Catherine Girouard, and Christian Gagn{\'{e}}.
\newblock {A two-step approach for mining patient treatment pathways in
  administrative healthcare databases}.
\newblock \emph{Artificial Intelligence in Medicine}, 87:\penalty0 34--48, 5
  2018.
\newblock ISSN 18732860.
\newblock \doi{10.1016/j.artmed.2018.03.004}.

\bibitem[Needleman and Wunsch(1970)]{Needleman}
Saul~B. Needleman and Christian~D. Wunsch.
\newblock A general method applicable to the search for similarities in the
  amino acid sequence of two proteins.
\newblock \emph{Journal of Molecular Biology}, 48\penalty0 (3):\penalty0
  443--453, 1970.
\newblock ISSN 0022-2836.
\newblock \doi{https://doi.org/10.1016/0022-2836(70)90057-4}.
\newblock URL
  \url{https://www.sciencedirect.com/science/article/pii/0022283670900574}.

\bibitem[Newman(2010)]{bookNewman}
Mark Newman.
\newblock \emph{{Networks: An Introduction}}.
\newblock Oxford University Press, 03 2010.
\newblock ISBN 9780199206650.
\newblock \doi{10.1093/acprof:oso/9780199206650.001.0001}.
\newblock URL \url{https://doi.org/10.1093/acprof:oso/9780199206650.001.0001}.

\bibitem[Page et~al.(1999)Page, Brin, Motwani, and Winograd]{pagerank}
Lawrence Page, Sergey Brin, Rajeev Motwani, and Terry Winograd.
\newblock The pagerank citation ranking: Bringing order to the web.
\newblock Technical report, Stanford InfoLab, 1999.

\bibitem[Peters et~al.(2013)Peters, Crespo, Lingras, and
  Weber]{ref_fuzzy_clustering}
Georg Peters, Fernando Crespo, Pawan Lingras, and Richard Weber.
\newblock Soft clustering – fuzzy and rough approaches and their extensions
  and derivatives.
\newblock \emph{International Journal of Approximate Reasoning}, 54\penalty0
  (2):\penalty0 307--322, 2013.
\newblock ISSN 0888-613X.
\newblock \doi{https://doi.org/10.1016/j.ijar.2012.10.003}.
\newblock URL
  \url{https://www.sciencedirect.com/science/article/pii/S0888613X12001739}.

\bibitem[Prodel et~al.(2018)Prodel, Augusto, Jouaneton, Lamarsalle, and
  Xie]{Prodel2018}
Martin Prodel, Vincent Augusto, Baptiste Jouaneton, Ludovic Lamarsalle, and
  Xiaolan Xie.
\newblock {Optimal Process Mining for Large and Complex Event Logs}.
\newblock \emph{IEEE Transactions on Automation Science and Engineering},
  15\penalty0 (3):\penalty0 1309--1325, 7 2018.
\newblock ISSN 15455955.
\newblock \doi{10.1109/TASE.2017.2784436}.

\bibitem[Prokofyeva and Zaytsev(2020)]{Prokofyeva2020a}
Elizaveta~S. Prokofyeva and Roman~D. Zaytsev.
\newblock {Clinical pathways analysis of patients in medical institutions based
  on hard and fuzzy clustering methods}.
\newblock \emph{Business Informatics}, 14\penalty0 (1):\penalty0 19--31, 2020.
\newblock ISSN 25878158.
\newblock \doi{10.17323/2587-814X.2020.1.19.31}.

\bibitem[Rebuge and Ferreira(2012)]{Rebuge2012}
{\'{A}}lvaro Rebuge and Diogo~R. Ferreira.
\newblock {Business process analysis in healthcare environments: A methodology
  based on process mining}.
\newblock \emph{Information Systems}, 37\penalty0 (2):\penalty0 99--116, 4
  2012.
\newblock ISSN 03064379.
\newblock \doi{10.1016/j.is.2011.01.003}.

\bibitem[Rinner et~al.(2018)Rinner, Helm, Dunkl, Kittler, and
  Rinderle-Ma]{Rinner2018}
Christoph Rinner, Emmanuel Helm, Reinhold Dunkl, Harald Kittler, and Stefanie
  Rinderle-Ma.
\newblock {Process mining and conformance checking of long running processes in
  the context of melanoma surveillance}.
\newblock \emph{International Journal of Environmental Research and Public
  Health}, 15\penalty0 (12), 12 2018.
\newblock ISSN 16604601.
\newblock \doi{10.3390/ijerph15122809}.

\bibitem[Rismanchian and Lee(2017)]{Rismanchian2017}
Farhood Rismanchian and Young~Hoon Lee.
\newblock {Process Mining–Based Method of Designing and Optimizing the
  Layouts of Emergency Departments in Hospitals}.
\newblock \emph{Health Environments Research and Design Journal}, 10\penalty0
  (4):\penalty0 105--120, 7 2017.
\newblock ISSN 19375867.
\newblock \doi{10.1177/1937586716674471}.

\bibitem[Rotondi et~al.(1997)Rotondi, Brindis, Cantees, DeRiso, Ilkin, Palmer,
  Gunnerson, and {David Watkins}]{Rotondi1997}
Armando~J. Rotondi, Charles Brindis, Kimberly~K. Cantees, Barbara~M. DeRiso,
  Hakan~M. Ilkin, Jeffrey~S. Palmer, Helena~B. Gunnerson, and W.~{David
  Watkins}.
\newblock Benchmarking the perioperative process. i. patient routing systems: A
  method for continual improvement of patient flow and resource utilization.
\newblock \emph{Journal of Clinical Anesthesia}, 9\penalty0 (2):\penalty0
  159--169, 1997.
\newblock ISSN 0952-8180.
\newblock \doi{https://doi.org/10.1016/S0952-8180(96)00242-5}.
\newblock URL
  \url{https://www.sciencedirect.com/science/article/pii/S0952818096002425}.

\bibitem[Sato et~al.(2020)Sato, Mantovani, Safanelli, Guesser, Nagel, Moro,
  Cabral, Scalabrin, Moro, and Santos]{Sato2020}
Denise~M.V. Sato, Let{\'{i}}cia~K Mantovani, Juliana Safanelli, Vanessa
  Guesser, Vivian Nagel, Carla~H.C. Moro, Norberto~L Cabral, Edson~E Scalabrin,
  Claudia Moro, and Eduardo~A.P. Santos.
\newblock {Ischemic stroke: Process perspective, clinical and profile
  characteristics, and external factors}.
\newblock \emph{Journal of Biomedical Informatics}, 111\penalty0 (1):\penalty0
  103582, 2020.
\newblock ISSN 15320464.
\newblock \doi{10.1016/j.jbi.2020.103582}.
\newblock URL \url{https://doi.org/10.1016/j.jbi.2020.103582}.

\bibitem[Senderovich et~al.(2016)Senderovich, Weidlich, Yedidsion, Gal,
  Mandelbaum, Kadish, and Bunnell]{Senderovich2016}
Arik Senderovich, Matthias Weidlich, Liron Yedidsion, Avigdor Gal, Avishai
  Mandelbaum, Sarah Kadish, and Craig~A. Bunnell.
\newblock {Conformance checking and performance improvement in scheduled
  processes: A queueing-network perspective}.
\newblock \emph{Information Systems}, 62:\penalty0 185--206, 12 2016.
\newblock ISSN 03064379.
\newblock \doi{10.1016/j.is.2016.01.002}.

\bibitem[Stefanini et~al.(2020)Stefanini, Aloini, Benevento, Dulmin, and
  Mininno]{Stefanini2020novo}
Alessandro Stefanini, Davide Aloini, Elisabetta Benevento, Riccardo Dulmin, and
  Valeria Mininno.
\newblock A process mining methodology for modeling unstructured processes.
\newblock \emph{Knowledge and Process Management}, 27\penalty0 (4):\penalty0
  294--310, 2020.
\newblock \doi{https://doi.org/10.1002/kpm.1649}.
\newblock URL \url{https://onlinelibrary.wiley.com/doi/abs/10.1002/kpm.1649}.

\bibitem[Su et~al.(2020)Su, Liu, Zheng, Zhou, and Zheng]{su2020survey}
Han Su, Shuncheng Liu, Bolong Zheng, Xiaofang Zhou, and Kai Zheng.
\newblock A survey of trajectory distance measures and performance evaluation.
\newblock \emph{The VLDB Journal}, 29:\penalty0 3--32, 2020.

\bibitem[Verbeek et~al.(2010)Verbeek, Buijs, van Dongen, and van~der
  Aalst]{ProM6}
H.~M.~W. Verbeek, Joos C. A.~M. Buijs, Boudewijn~F. van Dongen, and Wil M.~P.
  van~der Aalst.
\newblock Xes, xesame, and prom 6.
\newblock In Pnina Soffer and Erik Proper, editors, \emph{Information Systems
  Evolution - CAiSE Forum 2010, Hammamet, Tunisia, June 7-9, 2010, Selected
  Extended Papers}, volume~72 of \emph{Lecture Notes in Business Information
  Processing}, pages 60--75. Springer, 2010.
\newblock \doi{10.1007/978-3-642-17722-4\_5}.
\newblock URL \url{https://doi.org/10.1007/978-3-642-17722-4\_5}.

\bibitem[Vieira et~al.(2020)Vieira, Xavier, and
  Evsukoff]{survey_overlappig_communities}
Vin{\'\i}cius da~Fonseca Vieira, Carolina~Ribeiro Xavier, and
  Alexandre~Gon{\c{c}}alves Evsukoff.
\newblock A comparative study of overlapping community detection methods from
  the perspective of the structural properties.
\newblock \emph{Applied Network Science}, 5\penalty0 (1):\penalty0 1--42, 2020.

\bibitem[Vlachos et~al.(2002)Vlachos, Kollios, and
  Gunopulos]{vlachos2002discovering}
Michail Vlachos, George Kollios, and Dimitrios Gunopulos.
\newblock Discovering similar multidimensional trajectories.
\newblock In \emph{Proceedings 18th international conference on data
  engineering}, pages 673--684. IEEE, 2002.

\bibitem[Wasserman and Faust(1994)]{wasserman_faust_1994}
Stanley Wasserman and Katherine Faust.
\newblock \emph{Social Network Analysis: Methods and Applications}, pages
  200--201.
\newblock Structural Analysis in the Social Sciences. Cambridge University
  Press, 1994.
\newblock \doi{10.1017/CBO9780511815478}.

\bibitem[Wehmuth et~al.(2016)Wehmuth, Fleury, and Ziviani]{Wehmuth2016}
Klaus Wehmuth, {\'{E}}ric Fleury, and Artur Ziviani.
\newblock {On MultiAspect graphs}.
\newblock \emph{Theoretical Computer Science}, 651:\penalty0 50--61, 2016.
\newblock ISSN 03043975.
\newblock \doi{10.1016/j.tcs.2016.08.017}.

\bibitem[Zaballa et~al.(2020)Zaballa, P{\'{e}}rez, Inhiesto, Ayesta, and
  Lozano]{Zaballa2020}
Onintze Zaballa, Aritz P{\'{e}}rez, Elisa~G{\'{o}}mez Inhiesto,
  Teresa~Acaiturri Ayesta, and Jose~A. Lozano.
\newblock {Identifying common treatments from Electronic Health Records with
  missing information. An application to breast cancer}.
\newblock \emph{PLoS ONE}, 15\penalty0 (12 December), 12 2020.
\newblock ISSN 19326203.
\newblock \doi{10.1371/journal.pone.0244004}.

\bibitem[Zhang et~al.(2015{\natexlab{a}})Zhang, Padman, and Patel]{Zhang2015}
Yiye Zhang, Rema Padman, and Nirav Patel.
\newblock {Paving the COWpath: Learning and visualizing clinical pathways from
  electronic health record data}.
\newblock \emph{Journal of Biomedical Informatics}, 58:\penalty0 186--197, 12
  2015{\natexlab{a}}.
\newblock ISSN 15320464.
\newblock \doi{10.1016/j.jbi.2015.09.009}.

\bibitem[Zhang et~al.(2015{\natexlab{b}})Zhang, Padman, Wasserman, Patel,
  Teredesai, and Xie]{ZhangPadman2015}
Yiye Zhang, Rema Padman, Larry Wasserman, Nirav Patel, Pradip Teredesai, and
  Qizhi Xie.
\newblock {On clinical pathway discovery from electronic health record data}.
\newblock \emph{IEEE Intelligent Systems}, 30\penalty0 (1):\penalty0 70--75, 1
  2015{\natexlab{b}}.
\newblock ISSN 15411672.
\newblock \doi{10.1109/MIS.2015.14}.
\newblock URL \url{www.computer.org/intelligent
  http://ieeexplore.ieee.org/document/7030250/}.

\end{thebibliography}

% ----------------------------------------------

\newpage
\begin{appendices}

% ---------------------
\section{Criteria to identify patients who had a pregnancy}
\label{app:cids_sigtaps}
The patients who took part in the pregnancy group were those who had at least one record of the ICD-10 codes listed in Table~\ref{tab:lista_cids} or the medical procedure codes in Table~\ref{tab:lista_sigtaps}, between October/2014 and March/2015.

\begin{table}[!ht]
\caption{ICD-10 Codes used to identify pregnant patients.}
    \centering
\begin{tabular}[t]{|p{0.2\textwidth}<{\centering}|}
\hline
    ICD-10 Code \\  
    \hline
    O10 \\ 
    O11 \\ 
    O12 \\ 
    O13 \\ 
    O14 \\ 
    O15 \\ 
    O16 \\ 
    O20 \\ 
    O21 \\ 
    O22 \\ 
    O23 \\ 
    O24 \\ 
    O25 \\ 
    O26 \\ 
    O28 \\ 
    O29 \\ 
    O30 \\ 
    $\vdots$ \\ 
%\hline
\end{tabular}
\quad % if needed space between table
\begin{tabular}[t]{|p{0.2\textwidth}<{\centering}|}
 \hline
%     ICD-10 Code \\ \hline  
\\
    \rule{0pt}{4.5ex}$\vdots$ \\ 
    O31 \\ 
    O32 \\ 
    O33 \\ 
    O34 \\ 
    O35 \\ 
    O36 \\ 
    O40 \\ 
    O41 \\ 
    O42 \\ 
    O43 \\ 
    O44 \\ 
    O45 \\ 
    O46 \\ 
    O47 \\ 
    O48 \\ 
    $\vdots$ \\ 
% \hline
\end{tabular}
\quad % if needed space between table
\begin{tabular}[t]{|p{0.2\textwidth}<{\centering}|}
\hline
%     ICD-10 Code \\ \hline  
\\
    \rule{0pt}{4.5ex}$\vdots$ \\ 
    O60 \\ 
    O61 \\ 
    O62 \\ 
    O63 \\ 
    O64 \\ 
    O65 \\ 
    O66 \\ 
    O67 \\ 
    O68 \\ 
    O69 \\ 
    O70 \\ 
    O71 \\ 
    O72 \\ 
    O73 \\ 
    O74 \\ 
    $\vdots$ \\ 
% \hline
\end{tabular}
\quad % if needed space between table
\begin{tabular}[t]{|p{0.2\textwidth}<{\centering}|}
\hline
%     ICD-10 Code \\ \hline  
\\
    \rule{0pt}{4.5ex}$\vdots$ \\ 
    O75 \\ 
    O80 \\ 
    O81 \\ 
    O82 \\ 
    O83 \\ 
    O84 \\ 
    Z30 \\ 
    Z31 \\ 
    Z32 \\ 
    Z33 \\ 
    Z34 \\ 
    Z35 \\ 
    Z36 \\ 
    Z37 \\ 
    Z38 \\ 
    Z39 \\ 
    \\
\hline
\end{tabular}
\label{tab:lista_cids}
\end{table}

\begin{longtable}{|p{.15\textwidth}<{\centering}|p{.85\textwidth}|}
\caption{SIGTAP codes used to identify pregnant patients.} \label{tab:lista_sigtaps} \\
\hline
SIGTAP Code & Description (Portuguese)\\ \hline 
0801010012 & ADESAO A ASSISTENCIA PRE-NATAL - INCENTIVO PHPN (COMPONENTE I)\\ 
0211040010 & AMNIOSCOPIA\\ 
0801010047 & INCENTIVO AO REGISTRO CIVIL DE NASCIMENTO\\ 
04110100034 & PARTO CESARIANO\\ 
0303160055 & TRATAMENTO DAS MALFORMACOES E DEFORMIDADES CONGENITAS DO SISTEMA OSTEOMUSCULAR\\ 
0303100028 & TRATAMENTO DE ECLAMPSIA\\ 
0411020056 & TRATAMENTO DE OUTROS TRANSTORNOS MATERNOS RELACIONADOS PREDOMINANTEMENTE A GRAVIDEZ\\ 
0201010011 & AMNIOCENTESE\\ 
0417010028 & ANALGESIA OBSTETRICA P/ PARTO NORMAL\\ 
0417010010 & ANESTESIA OBSTETRICA P/ CESARIANA\\ 
0417010036 & ANESTESIA OBSTETRICA P/CESARIANA EM GESTACAO DE ALTO RISCO\\ 
0310010012 & ASSISTÊNCIA AO PARTO SEM DISTOCIA\\ 
0310010020 & ATENDIMENTO AO RECEM-NASCIDO NO MOMENTO DO NASCIMENTO\\ 
0302010033 & ATENDIMENTO FISIOTERAPÊUTICO EM PACIENTE NEONATO\\ 
0801010020 & CONCLUSAO DA ASSISTENCIA PRE-NATAL (INCENTIVO)\\ 
0301010110 & CONSULTA PRE-NATAL\\ 
0301010129 & CONSULTA PUERPERAL\\ 
0411020013 & CURETAGEM POS-ABORTAMENTO / PUERPERAL\\ 
0411010018 & DESCOLAMENTO MANUAL DE PLACENTA\\ 
0802010032 & DIARIA DE ACOMPANHANTE DE GESTANTE C/ PERNOITE\\ 
0411020021 & EMBRIOTOMIA\\ 
0202090167 & ESPECTROFOTOMETRIA NO LIQUIDO AMNIOTICO\\ 
0409060070 & ESVAZIAMENTO DE UTERO POS-ABORTO POR ASPIRACAO MANUAL INTRA-UTERINA (AMIU)\\ 
0411020030 & HISTERECTOMIA PUERPERAL\\ 
0801010039 & INCENTIVO AO PARTO - PHPN (COMPONENTE I)\\ 
0301100136 & ORDENHA MAMARIA\\ 
0411010042 & PARTO CESARIANO C/ LAQUEADURA TUBARIA\\ 
0411010026 & PARTO CESARIANO EM GESTACAO DE ALTO RISCO\\ 
0310010039 & PARTO NORMAL\\ 
0310010055 & PARTO NORMAL EM CENTRO DE PARTO NORMAL (CPN)\\ 
0310010047 & PARTO NORMAL EM GESTACAO DE ALTO RISCO\\ 
0301010145 & PRIMEIRA CONSULTA DE PEDIATRIA AO RECEM-NASCIDO\\ 
0304060178 & QUIMIOTERAPIA DE NEOPLASIA TROFOBLÁSTICA GESTACIONAL - BAIXO RISCO\\ 
0304060186 & QUIMIOTERAPIA DE NEOPLASIA TROFOBLÁSTICA GESTACIONAL - CORIOCARCINOMA DE BAIXO RISCO PERSISTENTE / ALTO RISCO / RECIDIVA\\ 
0304060208 & QUIMIOTERAPIA DE NEOPLASIA TROFOBLÁSTICA GESTACIONAL - CORIOMA / MOLA HIDATIFORME - PERSISTENTE / INVASIVA\\ 
0411010050 & REDUCAO MANUAL DE INVERSAO UTERINA AGUDA POS-PARTO\\ 
0411010069 & RESSUTURA DE EPISIORRAFIA POS-PARTO\\ 
0411010077 & SUTURA DE LACERACOES DE TRAJETO PELVICO (NO PARTO ANTES DA ADMISSAO)\\ 
0214010082 & TESTE RÁPIDO PARA SÍFILIS EM GESTANTE\\ 
0306020157 & TRANSFUSAO FETAL INTRA-UTERINA\\ 
0411020048 & TRATAMENTO CIRURGICO DE GRAVIDEZ ECTOPICA\\ 
0411010085 & TRATAMENTO CIRURGICO DE INVERSAO UTERINA AGUDA POS PARTO\\ 
0303100010 & TRATAMENTO DE COMPLICACOES RELACIONADAS PREDOMINANTEMENTE AO PUERPERIO\\ 
0303100036 & TRATAMENTO DE EDEMA, PROTEINURIA E TRANSTORNOS HIPERTENSIVOS NA GRAVIDEZ PARTO E PUERPERIO\\ 
0303100044 & TRATAMENTO DE INTERCORRENCIAS CLINICAS NA GRAVIDEZ\\ 
0303100052 & TRATAMENTO DE MOLA HIDATIFORME\\ 
0205020143 & ULTRA-SONOGRAFIA OBSTETRICA\\ 
0205020151 & ULTRASSONOGRAFIA OBSTETRICA C/ DOPPLER COLORIDO E PULSADO\\ 
0205020178 & ULTRASSONOGRAFIA TRANSFONTANELA\\ 
0202031179 & VDRL P/ DETECCAO DE SIFILIS EM GESTANTE \\ \hline
\end{longtable}

\clearpage
% ----------------

\section{Criteria to identify diabetic patients}
\label{app:cids_diabetes}

The patients who took part in the diabetes group were those who had at least one record of the ICD-10 codes listed in Table~\ref{tab:lista_cids_diabetes}.

\begin{table}[!ht]
    \centering
    \caption{ICD-10 Codes used to identify diabetic patients.}
    \label{tab:lista_cids_diabetes}
    \begin{tabular}{|c|}
    \hline
    ICD-10 Code \\ \hline
        E10 \\ 
        E11 \\ 
        E12 \\ 
        E13 \\ 
        E14 \\ 
        N083 \\ 
        N251 \\ 
        O240 \\ 
        O241 \\ 
        O242 \\ 
        O243 \\ 
        Z131 \\ 
        Z833 \\ \hline
    \end{tabular}
    
\end{table}

\clearpage

% --------------------
% \section{Example -- patient pathway similarity}
% We illustrate the use of the similarity measure with the synthetic pathways presented in Figure \ref{fig:example_pathway_distance}, where ${x_A,x_B,x_C,x_D}$ is the set of activity tuples and the numbers above the arrows indicate the time interval in days between the activity tuples. %$\{x_{i}   : i \in \{A,B,C,D\}$\} 

% \begin{figure}[!ht]
%     \centering
%     %\includegraphics{}
%     \begin{center}
%     \leavevmode
%     \resizebox{0.4\textwidth}{!}{
%         \xymatrix@C=1.5em@R=1em{
%         Pathway\; P : & x_{A} \ar[r]^{2} & x_{B} \ar[r]^{3} & x_{C} \\
%         Pathway\; P^{\prime} : & x_{A} \ar[r]^{5} & x_{C} }
%     }
%     \end{center}
%     \caption{Caption}
%     \label{fig:example_pathway_distance}
% \end{figure}

% For simplicity, we define $Dist_A$ as in Equation \ref{eq:distA_example_pathway_distance} and set $\delta = 0$; in other words, we will only align two elements of pathways $P$ and $P^{\prime}$ if the activity tuples are identical.

% \begin{equation*}
% \label{eq:distA_example_pathway_distance}
%     Dist_A(x_i,x_j) = \begin{cases}
%         1, & \quad i \neq j \\
%         0, & \quad i = j
% \end{cases} \quad \forall i,j \in \{A,B,C,D\}.
% \end{equation*}

% \begin{figure}[!ht]
%     \centering\includegraphics[width=\textwidth]{fig/exemplo_medida_similaridade.jpg}
%     \caption{Caption}
%     \label{fig:example_pathway_distance_diagram}
% \end{figure}
% \clearpage
% -------------------
\end{appendices}

\end{document}